\documentclass[fleqn,usenatbib]{mnras}

\usepackage{newtxtext,newtxmath}

\usepackage[T1]{fontenc}
\usepackage{ae,aecompl}

\usepackage{graphicx}	
\usepackage{amsmath}	

\title[Low-frequency survey of SPT cluster sample]{A low-frequency radio halo survey of the South Pole Telescope SZ-selected clusters with the GMRT}

\author[R. Raja et al.]{
Ramij Raja,$^{1}$\thanks{E-mail: phd1601121008@iiti.ac.in}
Majidul Rahaman,$^{1}$
Abhirup Datta,$^{1}$
Reinout J. van Weeren$^{2}$
\newauthor
Huib T. Intema,$^{2,3}$
and Surajit Paul$^{4}$
\\
$^{1}$Discipline of Astronomy, Astrophysics and Space Engineering, Indian Institute of Technology Indore, Simrol, 453552, India\\
$^{2}$Leiden Observatory, Leiden University, PO Box 9513, 2300 RA Leiden, The Netherlands \\
$^{3}$International Centre for Radio Astronomy Research, Curtin University, GPO Box U1987, Perth, WA 6845, Australia\\
$^{4}$Department of Physics, Savitribai Phule Pune University, Pune 411007, India\\
}

\date{Accepted XXX. Received YYY; in original form ZZZ}

\pubyear{2015}

\begin{document}
\label{firstpage}
\pagerange{\pageref{firstpage}--\pageref{lastpage}}
\maketitle

\begin{abstract}
The presence of non-thermal electrons and large scale magnetic fields in the intra-cluster medium (ICM) is known through the detection of mega-parsec (Mpc) scale diffuse radio synchrotron emission. Although a significant amount of progress in finding new diffuse radio sources has happened in the last decade, most of the investigation has been constrained towards massive low-redshift clusters. In this work, we explore clusters with redshift $z>0.3$ in search of diffuse radio emission, at 325 MHz with the Giant Metrewave Radio Telescope (GMRT). This campaign has resulted in the discovery of 2 new radio halos (SPT-CL J0013-4906 and SPT-CL J0304-4401) along with 2 other detections (SPT-CL J2031-4037 and SPT-CL J2248-4431), previously reported (at 325 MHz) in the literature. In addition, we detect a halo candidate in 1 cluster in our sample, and upper limits for halos are placed in 8 clusters where no diffuse emission is detected. In the $P_{1.4} - L_\mathrm{X}$ plane, the detected halos follow the observed correlation, whereas the upper limits lie above the correlation line, indicating the possibility of future detection with sensitive observations. 

\end{abstract}




\section{Introduction} \label{sec:intro}
The diffuse hot ($10^7-10^8$ K) plasma known as intracluster medium (ICM) situated within the cluster gravitational potential makes up $\sim15\%$ of the cluster mass, which emits in the soft X-ray band via thermal Bremsstrahlung. Clusters grow via accretion, small and large mergers of sub-clusters, or galaxy groups. The major merger events are some of the most energetic events since the Big Bang, releasing up to $10^{64}$ ergs of energy within a few Gyr of timescale. This enormous amount of energy is dissipated into the ICM primarily via weak shocks and turbulence (e.g., \citealt{Paul2011ApJ...726...17P}). These shocks and turbulence re-accelerate \textit{in situ} relativistic electrons in the ICM, which emit synchrotron radiation in the presence of large scale magnetic fields, and forms different diffuse structures depending upon the merger type (e.g., major/minor or on-axis/off-axis). The diffuse radio sources found in the galaxy clusters are typically divided into three different categories: (1) halos, (2) relics, and (3) minihalos (see  \citealt{Feretti2012A&ARv..20...54F,Brunetti2014IJMPD..2330007B,vanWeeren2019SSRv..215...16V} for review).

Radio halos are Mpc scale radio objects with regular morphology that roughly traces the cluster ICM visible in X-ray, and does not have an optical counterpart. They are found primarily in the merging clusters. It has unpolarised emission with a typical spectral index\footnote{The synchrotron emission spectral index $\alpha$ is defined as $S_{\nu}=\nu^{\alpha}$, where $S_{\nu}$ is the flux density at frequency $\nu$.} of $\alpha\sim-1.2$. The prototype of this class is the halo found in the Coma cluster \citep{Large1959Natur.183.1663L}. The primary source of synchrotron emitting electrons is generated through re-acceleration of \textit{in situ} relativistic electrons via merger driven turbulence \citep{Brunetti2001MNRAS.320..365B,Petrosian2001ApJ...557..560P}. Also, the contribution of the secondary electrons generated via proton-proton collision in the ICM was proposed by e.g., \citet{Dennison1980ApJ...239L..93D,Dolag2000A&A...362..151D,Brunetti2011MNRAS.410..127B}. However, results from recent gamma-ray observations have excluded pure hadronic origin of radio halo and limited their contribution in providing seed electrons for further re-acceleration \citep{Ackermann2014ApJ...787...18A,Brunetti2017MNRAS.472.1506B}. Besides, there is a subclass of halos that are predicted in the turbulent re-acceleration mechanism corresponding to the less-energetic merger events, which has a very steep spectral index ($\alpha<-1.5$; e.g., \citealt{Brunetti2008Natur.455..944B}).

Radio relics are Mpc scale radio objects with elongated morphology, which are found in the periphery of clusters. They trace shock waves found in X-ray observations (e.g., \citealt{Finoguenov2010ApJ...715.1143F}). They also are found in merging clusters and have a high degree of polarization ($\gtrsim20\%$ at GHz frequencies; \citealt{Enblin1998A&A...332..395E}) with a typical spectral index of $-1<\alpha<-1.5$. The prototype of this class is the relic in the CIZAJ2242.8+5301 cluster \citep{vanWeeren2010Sci...330..347V}. The relativistic electrons emitting in these diffuse sources are re-accelerated at the merging shock fronts via DSA (Diffusive Shock Acceleration) mechanism (e.g., \citealt{Enblin1998A&A...332..395E}). However, this simple DSA mechanism faces the problem of requiring unrealistic re-acceleration efficiency to supply necessary relativistic electrons from the thermal pool \citep{Macario2011ApJ...728...82M,Eckert2016MNRAS.461.1302E,vanWeeren2016ApJ...818..204V}. Nevertheless, this problem is resolved if the seed electrons are provided by nearby radio galaxy, and indeed some observational evidence supports this scenario (e.g., \citealt{Bonafede2014ApJ...785....1B,Shimwell2015MNRAS.449.1486S,vanWeeren2017ApJ...835..197V}). 

Radio minihalos are a smaller version of radio halos with similar morphology and location in the clusters. However, unlike radio halos, they are found only in relaxed cool-core clusters. The observed radio emission from these sources are unpolarised and has a typical integrated spectral index of about $-1.1$. The prototype of this class is the minihalo found in the Perseus cluster (e.g., \citealt{Pedlar1990MNRAS.246..477P,Gendron-Marsolais2017MNRAS.469.3872G}). A proposed origin of synchrotron emitting electrons in the minihalos is the \lq\lq sloshing\rq\rq\ driven turbulence generated by minor merger events in the cool-core clusters \citep{Fujita2004JKAS...37..571F,Mazzotta2008ApJ...675L...9M,ZuHone2013ApJ...762...78Z}. However, similar to radio halos, the role of secondary mechanism in supplying relativistic electrons were also proposed for minihalos as well (e.g., \citealt{Pfrommer2004A&A...413...17P,Fujita2007ApJ...663L..61F,Keshet2010ApJ...722..737K}).

Apart from these broad classes, recent sensitive low-frequency observations have discovered some other complex, intermediate diffuse radio sources, which are challenging our understanding of the astrophysical processes in the galaxy clusters (see \citealt{vanWeeren2019SSRv..215...16V} for review).

A considerable amount of progress has happened in the last decade in detecting new radio halos, by almost doubling the numbers that were previously discovered. However, a systematic search of radio halos was carried out so far only by  \citet{Venturi2007A&A...463..937V,Venturi2008A&A...484..327V} and \citet{Kale2013A&A...557A..99K,Kale2015A&A...579A..92K} in the GMRT Radio Halo Survey (GRHS) and Extended-GRHS (EGRHS), respectively. Apart from increasing the number of known halos, these surveys have shaped our statistical understanding of the radio halos \citep{Cassano2013ApJ...777..141C,Kale2015A&A...579A..92K}. Nevertheless, the domain explored in the galaxy cluster distribution in these surveys is limited. Both of these survey and other targeted observations were aimed towards high-mass low-redshift clusters, as a potential source of radio halo. In this work, we have complemented the previous surveys by exploring higher redshift clusters in our sample. Recently, similar work has been presented by \citet{Knowles2019MNRAS.486.1332K} where they have studied SZ-selected high mass ($> 5 \times 10^{14} M_{\odot}$) clusters beyond redshift 0.5. Furthermore, a study of high redshift ($z>0.3$) X-ray selected clusters were also performed by \citet{Giovannini2020A&A...640A.108G}.

In this paper, we present radio observational results of 15 SZ-Xray selected clusters and compared their statistical properties with the literature halos. In Sect. \ref{sec:sample}, we have presented our cluster sample. The radio observations and data reduction method are described in Sect. \ref{sec:obs}. The results are presented in Sect. \ref{sec:results} with confirmed halos, candidate halos and halo upper limits are presented separately in subsections. The implications of the result are discussed in Sect. \ref{sec:discuss}, and finally, our conclusions are presented in Sect. \ref{sec:conclude}.

In this work, we have adopted a $\Lambda$CDM cosmology with $H_0 = 70$ km s$^{-1}$ Mpc$^{-1}$, $\Omega_{\mathrm{m}} = 0.3$ and $\Omega_\Lambda = 0.7$.

\section{Cluster Sample} \label{sec:sample}
The sample presented here consists of 18 SPT-SZ survey clusters, which are selected from a sample of 83 clusters of the same, presented by \citet{McDonald2013ApJ...774...23M}. All these clusters have \textit{Chandra} X-ray observations with spectroscopic redshifts $z > 0.3$. The GMRT observations of the sample were done in two phases. In the first phase, 13 clusters were observed that have (1) central temperature $T_{\mathrm{central}} > 7$ keV, (2) central cooling time $t_{\mathrm{cool,0}} > 3$ Gyr, (3) surface brightness concentration parameter \citep{Santos2008A&A...483...35S} $c_{\mathrm{SB}} < 0.1$ to avoid cool core clusters and (4) a disturbed morphology in the \textit{Chandra} X-ray map. With the above criteria, a sample of clusters with different mass and disturbed morphology were selected which have a relatively high probability of detecting diffuse radio emission. In the second phase, the requirements of the central temperature $T_{\mathrm{central}} > 7$ keV, and the central cooling time $t_{\mathrm{cool,0}} > 3$ Gyr were relaxed to include some lower mass clusters as well in the proposed sample of 14 clusters. However, due to less time allocation, only 5 massive clusters were observed. Nevertheless, this relaxation did include some intermediate (SPT-CL J0040-4407, SPT-CL J2031-4037, SPT-CL J2248-4431) and a relaxed (SPT-CL J0304-4921) cluster. Apart from that, all these clusters have declination $> -50^{\circ}30\arcmin$ to allow up to 3-hour of continuous observation with the GMRT.

From these 18 clusters, 3 clusters had bad radio observations (explained in the Sect. \ref{sec:obs}), and are excluded from this study. The final sample, hereafter referred to as GMRT SPT-SZ Radio Halo Survey or GSRHS, consists of 15 clusters, spanning a redshift range of $\sim 0.3 < z < 0.83$ and a mass range of $\sim 4.5 \times 10^{14} < M_{500} < 18 \times 10^{14} M_{\odot}$ with the SPT detection significance range of $\sim 7 < \xi < 42$. 
This sample is presented in Table \ref{tab:sample} along with their global properties.

\begin{table*}
    \caption{Global cluster properties of the sample}
    \label{tab:sample}
    \scalebox{0.95}{
    \begin{tabular}{lcccccccccc}
    \hline
    \noalign{\smallskip}
    Name & $\mathrm{RA_{J2000}}$ $\mathrm{DEC_{J2000}}$ & $z$ & $M_\mathrm{{500}}$ & $L_\mathrm{X}$ & $T_\mathrm{central}$ & $c_\mathrm{SB}$ & $w$ & $t_\mathrm{cool}$ & Morph. & kpc/$\arcsec$ \\
    \noalign{\smallskip}
    SPT-CL   & hh mm ss\  $^\circ\ \arcmin\ \arcsec$ & & ($10^{14}\ M_{\odot}$) & ($10^{44}$\ erg s$^{-1}$) & (keV) & & ($10^{-2}$) & (Gyr) & & \\
    \hline
    \noalign{\smallskip}
    J0013-4906 & 00 13 19.44  -49 06 57.60 & 0.406 & $7.08 \pm 1.15$ & $5.6 \pm 0.4$ & 8.0 & 0.16 & 2.9 & 6.83 & M & 5.43 \\
    \noalign{\smallskip}
    J0014-4952 & 00 14 45.60  -49 52 51.60 & 0.752 & $5.31 \pm 0.92$ & $6.8 \pm 0.5$ & 7.4 & 0.13 & 10.1 & 11.44 & M & 7.35 \\
    \noalign{\smallskip}
    J0040-4407$^*$ & 00 40 46.63  -44 07 58.09 & 0.350 & $10.18 \pm 1.32$ & $5.9 \pm 0.4$ & 7.5 & $-$ & 0.9 & 2.34 & I & 4.94 \\
    \noalign{\smallskip}
    J0123-4821 & 01 23 11.04  -48 21 28.80 & 0.655 & $4.46 \pm 0.87$ & $2.9 \pm 0.2$ & 7.3 & 0.16 & 3.9 & 9.77 & M & 6.96 \\
    \noalign{\smallskip}
    J0142-5032 & 01 42 11.04  -50 32 24.00 & 0.6793 & $5.75 \pm 0.95$ & $4.3 \pm 0.5$ & 8.5 & 0.12 & 2.8 & 10.03 & M & 7.07 \\
    \noalign{\smallskip}
    J0212-4657 & 02 12 25.92  -46 57 0.00 & 0.655 & $5.88 \pm 0.98$ & $5.4 \pm 0.6$ & 8.2 & 0.15 & 7.5 & 12.14 & M & 6.96 \\
    \noalign{\smallskip}
    J0304-4401 & 03 04 20.29  -44 02 27.80 & 0.458 & $8.55 \pm 1.32$ & $8.7 \pm 0.4$ & 10.1 & 0.12 & 4.4 & 14.72 & M & 5.82 \\
    \noalign{\smallskip}
    J0304-4921 & 03 04 16.00  -49 21 26.30 & 0.392 & $7.57 \pm 1.2$ & $6.0 \pm 0.3$ & 4.0 & 0.33 & 1.0 & 0.53 & R & 5.31 \\
    \noalign{\smallskip}
    J0307-5042 & 03 07 50.64  -50 42 18.00 & 0.550 & $5.26 \pm 0.93$ & $4.2 \pm 0.3$ & 7.2 & 0.21 & 1.5 & 6.31 & M & 6.43 \\
    \noalign{\smallskip}
    J0348-4515 & 03 48 15.95  -45 14 42.75 & 0.358 & $6.17 \pm 1.03$ & $3.2 \pm 0.3$ & 2.4 & 0.16 & 1.6 & 1.39 & M & 5.01 \\
    \noalign{\smallskip}
    J0411-4819 & 04 11 9.40  -48 18 10.80 & 0.424 & $8.18 \pm 1.27$ & $7.4 \pm 0.3$ & 7.8 & 0.22 & 5.7 & 9.18 & M & 5.58 \\
    \noalign{\smallskip}
    J0449-4901 & 04 49 4.00  -49 01 39.00 & 0.792 & $4.57 \pm 0.86$ & $3.9 \pm 0.5$ & 9.8 & 0.13 & 4.6 & 11.14 & M & 7.5 \\
    \noalign{\smallskip}
    J0456-5116 & 04 56 28.09  -51 16 35.00 & 0.562 & $5.09 \pm 0.89$ & $3.5 \pm 0.2$ & 10.8 & 0.19 & 2.3 & 6.32 & M & 6.48 \\
    \noalign{\smallskip}
    J2031-4037 & 20 31 51.49  -40 37 14.02 & 0.3416 & $9.83 \pm 1.15$ & $6.8 \pm 0.3$ & 12.2 & 0.22 & 1.7 & 3.43 & I & 4.87 \\
    \noalign{\smallskip}
    J2248-4431 & 22 48 54.30  -44 31 7.00 & 0.351 & $17.97 \pm 2.18$ & $25.1 \pm 0.3$ & 13.0 & 0.23 & 0.6 & 1.79 & I & 4.95 \\
    \noalign{\smallskip}
    J2258-4044$^*$ & 22 58 49.44  -40 44 24.00 & 0.8971 & $5.88 \pm 0.95$ & $5.5 \pm 0.5$ & 7.9 & $-$ & 3.6 & 5.98 & M & 7.8 \\
    \noalign{\smallskip}
    J2301-4023 & 23 01 53.04  -40 23 20.40 & 0.8349 & $4.81 \pm 0.86$ & $4.8 \pm 0.4$ & 9.6 & 0.3 & 2.8 & 3.0 & M & 7.62 \\
    \noalign{\smallskip}
    J2325-4111$^*$ & 23 25 8.25  -41 12 42.59 & 0.358 & $7.55 \pm 1.2$ & $4.8 \pm 0.3$ & 10.4 & $-$ & 4.0 & 13.06 & M & 5.01 \\
    \hline
    \end{tabular}}
\\\flushleft{\textit{Note.} The columns are 1. Cluster name, 2. Right Ascension, Declination, 3. Redshift, 4. Mass within $R_{500}$, 5. X-ray luminosity in the energy range $0.1-2.4$ keV, 6. Central temperature, 7. X-ray surface brightness concentration parameter \citep{Santos2008A&A...483...35S}, 8. Centroid shift \citep{Mohr1993ApJ...413..492M}, 9. Central cooling time, 10. Morphology status (M = Merging or Non-cool-core, R = Relaxed or cool-core and I = intermediate or Weak cool-core), 11. Linear scale at respective redshift. The $M_\mathrm{{500}}$ information of the clusters are taken from \citet{Bleem2015ApJS..216...27B}. The $T_\mathrm{central}$ and $t_\mathrm{cool}$ values are taken from \citet{McDonald2013ApJ...774...23M}. The $c_\mathrm{SB}$ values are derived in this work. The $w$ values are taken from \citet{Nurgaliev2017ApJ...841...5N}. The clusters with $^*$ sign had bad radio data and are not included in this study.}
\end{table*}

\section{Radio observations, Data reduction and Imaging} \label{sec:obs}
The observations of the whole sample (18 clusters) were carried out in two GMRT observation cycles (Project code: 26\_024 and 27\_026) with a total of about 98 hrs observing time, including overheads. All observations were made in dual polarisation mode at 325 MHz with 32 MHz bandwidth divided into 256 spectral channels. 

The data reduction was done using SPAM pipeline \citep{Intema2017A&A...598A..78I}, which performs Radio Frequency Interference (RFI) mitigation, direction-dependent calibration, and ionospheric modelling \citep{Intema2009A&A...501.1185I}. \citet{Scaife2012MNRAS.423L..30S} scale was used to set flux densities of the calibrators. We adopted a flux density uncertainty of 10\% \citep{Chandra2004ApJ...612..974C}, which were quadratically added to the respective map noise of the images.

For low-resolution image, we tried with different values of robust parameter \citep{Briggs1995}, and uv-taper, and found robust = 0 with uv-taper at 7 k$\lambda$ to be the most suitable for most of the clusters, with the best signal-to-noise to the large scale structure. 

For each cluster, the high-resolution image of the unresolved point sources in the cluster region was made to distinguish from extended emission. To do that, we have excluded baselines shorter than 3k$\lambda$\footnote{For exceptions, see Table \ref{tab:Image_info}} (to get rid of the diffuse emission) and chose robust = -1 (to get more uniform UV-coverage, resulting in higher resolution) during the imaging. The resultant images are shown as blue/red contours overlaid on optical image for respective clusters. We estimated the flux densities of these radio galaxies using PyBDSF \citep{Mohan2015ascl.soft02007M} and were subtracted from the total radio emission of the respective cluster to estimate the diffuse radio flux density. The flux densities of these radio galaxies are presented in Table \ref{tab:Image_info}.

Due to the presence of bright point source in the field of view whose sidelobes contaminated the target region, 3 out of 18 cluster data were unusable. These clusters are namely SPT-CL J0040-4407, SPT-CL J2258-4044 and SPT-CL J2325-4111.

\begin{table}
    \caption{Imaging information}
    \label{tab:Image_info}
    \scalebox{0.92}{
    \begin{tabular}{ccccc}
    \hline
    \noalign{\smallskip}
    Name & uv-cut & uv-taper & Beam & $\sigma_{\mathrm{rms}}$ \\
    \noalign{\smallskip}
    SPT-CL & (k$\lambda$) & (k$\lambda$) & ($\arcsec \times \arcsec$, $^\circ$) & ($\mu$Jy beam$^{-1}$) \\
    \hline
    J0013-4906 & $-$ & $-$ & $32.1 \times 9.1,-2.3$ & 80 \\
     & $>3$ & $-$ & $28.3 \times 6.3,-2.6$ & 100 \\
    \hline
    J0014-4952 & $-$ & 7 & $31.0 \times 13.4,1.5$ & 100 \\
     & $>3$ & $-$ & $22.3 \times 5.0,-0.4$ & 80 \\
    \hline
    J0123-4821 & $-$ & 7 & $35.4 \times 13.0,0.1$ & 100 \\
     & $>3$ & $-$ & $24.2 \times 6.3,-3.1$ & 100 \\
    \hline
    J0142-5032 & $-$ & 7 & $36.3 \times 13.3,5.3$ & 100 \\
     & $>3$ & $-$ & $24.7 \times 6.4,2.4$ & 80 \\
    \hline
    J0212-4657 & $-$ & 7 & $36.4 \times 13.3,10.2$ & 180 \\
     & $>3$ & $-$ & $26.1 \times 6.4,4.1$ & 130 \\
    \hline
    J0304-4401 & $-$ & 7 & $35.0 \times 25.0,0.0$ & 180 \\
     & $>3$ & $-$ & $18.6 \times 6.6,10.0$ & 80 \\
    \hline
    J0304-4921 & $-$ & 7 & $34.2\times14.0,12.7$ & 150 \\
     & $>3$ & $-$ & $25.3\times7.0,14.8$ & 150 \\
    \hline
    J0307-5042 & $-$ & 7 & $34.4\times13.1,7.8$ & 80 \\
     & $>3$ & $-$ & $25.2\times6.3,1.7$ & 80 \\
    \hline
    J0348-4515 & $-$ & $-$ & $35.9\times11.5,33.5$ & 150 \\
     & $>3$ & $-$ & $23.8\times7.7,28.5$ & 100 \\
    \hline
    J0411-4819 & $-$ & 7 & $43.9\times13.2,15.9$ & 200 \\
     & $>3$ & $-$ & $31.5\times6.3,15.2$ & 200 \\
    \hline
    J0449-4901 & $-$ & 7 & $36.7\times12.9,6.2$ & 130 \\
     & $>3$ & $-$ & $23.4\times5.0,5.3$ & 200 \\
    \hline
    J0456-5116 & $-$ & 7 & $42.3\times12.7,7.1$ & 150 \\
     & $>3$ & $-$ & $27.3\times6.4,3.1$ & 150 \\
    \hline
    J2031-4037 & $-$ & $-$ & $21.5\times10.1,-3.5$ & 60 \\
     & $>5$ & $-$ & $14.9\times4.9,2.0$ & 120 \\
    \hline
    J2248-4431 & $-$ & $-$ & $23.0\times23.0,0.0$ & 100 \\
     & $>1$ & $-$ & $18.1\times5.1,6.8$ & 100 \\
    \hline
    J2301-4023 & $-$ & 7 & $33.5\times13.7,6.2$ & 180 \\
     & $>3$ & $-$ & $22.9\times6.5,2.8$ & 100 \\
    \hline
    \end{tabular}}
\\\flushleft{\textit{Note.} For each cluster, the first row corresponds to the low-resolution image shown in black contours, and the second row corresponds to the high-resolution image shown in blue/red contours in the respective figures. The \citet{Briggs1995} robust parameter used for low and high-resolution images are 0 and -1, respectively (except for SPT-CL J2031-4037, where robust=0.5).}
\end{table}

\section{Results} \label{sec:results}
The results of our observations in reporting the presence and absence of diffuse radio emission in the clusters are presented in the following sections. A summary of the properties of the discrete and diffuse radio sources are presented in Table \ref{tab:flux}.

\subsection{Diffuse radio emission} \label{subsec:diffuse emission}
\subsubsection{SPT-CL J0013-4906}
The SPT-CL J0013-4906 \citep{McDonald2013ApJ...774...23M} is massive ($M_{500}$ = $(7.1 \pm 1.1) \times 10^{14}$ $M_{\odot}$; \citealt{Bleem2015ApJS..216...27B}) cluster situated at the redshift $z = 0.406$ \citep{Bleem2015ApJS..216...27B}. The \textit{Chandra} X-ray luminosity of this cluster is $L_{[0.1-2.4\ \mathrm{keV}]} = (5.6 \pm 0.4) \times 10^{44}$ erg s$^{-1}$. The central temperature of the cluster and the morphology parameters suggest that it is a disturbed, non-cool-core cluster (Table \ref{tab:sample}).

\begin{figure*}
    \begin{tabular}{cc}
    \includegraphics[width=\columnwidth]{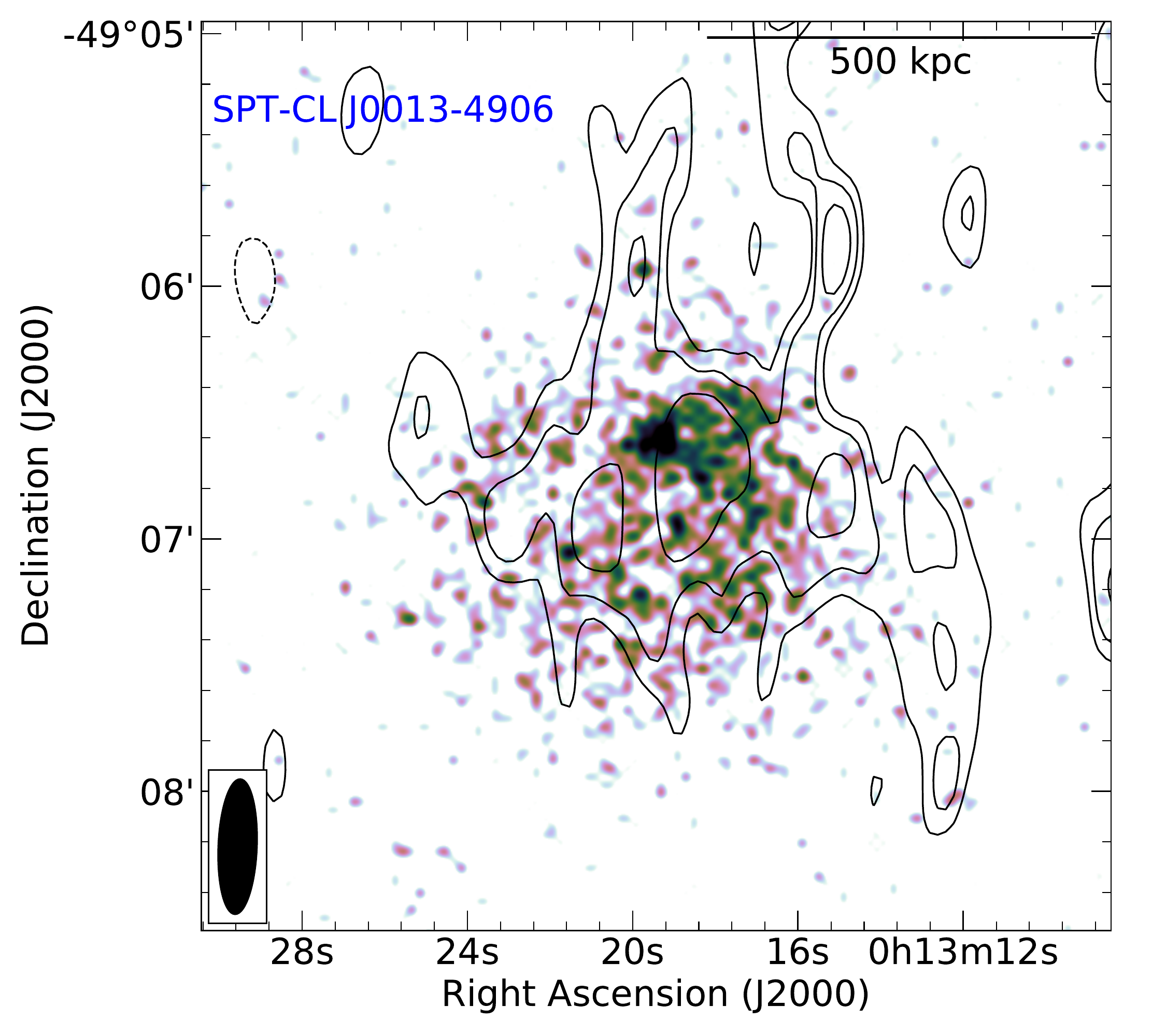} &
    \includegraphics[width=\columnwidth]{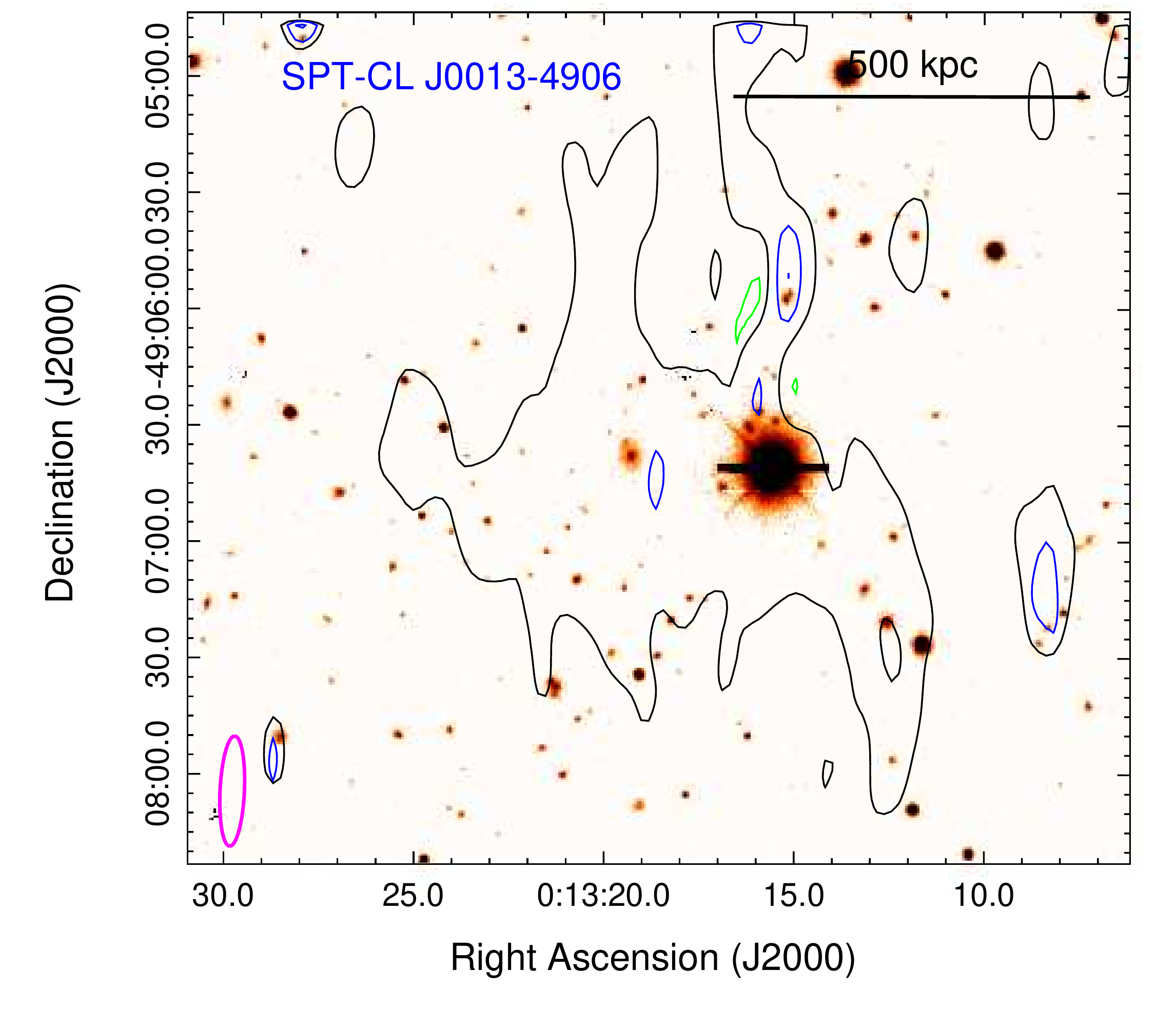} 
    \end{tabular}
    \caption{\textit{Left}: \textit{Chandra} X-ray image overlaid with 325 MHz radio contours of the SPT-CL J0013-4906 cluster. The restoring beam of the radio image is $32\arcsec \times 9\arcsec$, PA $-3^{\circ}$, indicated at the bottom left corner. The black contours are drawn at levels $[-1, 1, \sqrt{2}, 2, 4,...] \times 3\sigma_{\mathrm{rms}}$ with $\sigma_{\mathrm{rms}}$ = 80 $\mu$Jy beam$^{-1}$. Negative contours are indicated with dotted lines. 
    \textit{Right}: DECam optical image is overlaid with 325 MHz high-resolution image contours excluding baselines shorter than 3k$\lambda$ (blue), and are drawn at levels $[-1, 1, 2, 4, 8,...] \times 3\sigma_{\mathrm{rms}}$ with $\sigma_{\mathrm{rms}}$ = 100 $\mu$Jy beam$^{-1}$. Negative contours are indicated in green. The restoring beam of the radio image is $28\arcsec \times 6\arcsec$, PA $-3^{\circ}$, indicated at the bottom left corner in magenta. The 3$\sigma_{\mathrm{rms}}$ contour of the \textit{left panel} is shown in black to show the extent of the diffuse emission.}
    \label{fig:0013-4096}
\end{figure*}

The 325 MHz images are presented in Fig. \ref{fig:0013-4096}. 
The diffuse radio emission covers almost all of the cluster region visible in the X-ray image (Fig. \ref{fig:0013-4096} \textit{left panel}). The size of the diffuse emission is about $2.4\arcmin \times 3.0\arcmin$ or $0.8 \times 1$ Mpc (East-West $\times$ North-South or E-W $\times$ N-S). The Dark Energy Camera (DECam) optical image shows a BCG (Brightest Cluster Galaxy) at the position of the X-ray peak with a spectroscopic redshift of $z = 0.4099$ \citep{Bayliss2016ApJS..227....3B}.
No obvious radio counterpart of this galaxy is visible in the 325 MHz high-resolution image (blue contours in Fig. \ref{fig:0013-4096} \textit{right panel}). However, the $3\sigma$ significance radio emission near the BCG may be associated with this source.
The flux density of the diffuse radio emission was found to be $11.01 \pm 1.19$ mJy at 325 MHz.
Considering the cluster morphology and the diffuse radio structure, we classify this extended emission as a radio halo.  

\subsubsection{SPT-CL J0304-4401}
The SPT-CL J0304-4401 was first discovered by \citet{Williamson2011ApJ...738..139W} in the SPT-SZ survey. It is a massive ($M_{500}$ = $(8.5 \pm 1.3) \times 10^{14}$ $M_{\odot}$; \citealt{Bleem2015ApJS..216...27B}) cluster situated at the redshift $z = 0.458$ \citep{Bleem2015ApJS..216...27B}. The \textit{Chandra} X-ray image of this cluster shows a disturbed morphology with three distinct ICM gas clumps, possibly due to recent or ongoing merging event. The X-ray luminosity of this cluster was found to be $L_{[0.1-2.4\ \mathrm{keV}]} = (8.7 \pm 0.4) \times 10^{44}$ erg s$^{-1}$. The central temperature and the morphology parameters also suggest that it is a disturbed, non-cool-core cluster (Table \ref{tab:sample}).

\begin{figure*}
    \begin{tabular}{cc}
    \includegraphics[width=\columnwidth]{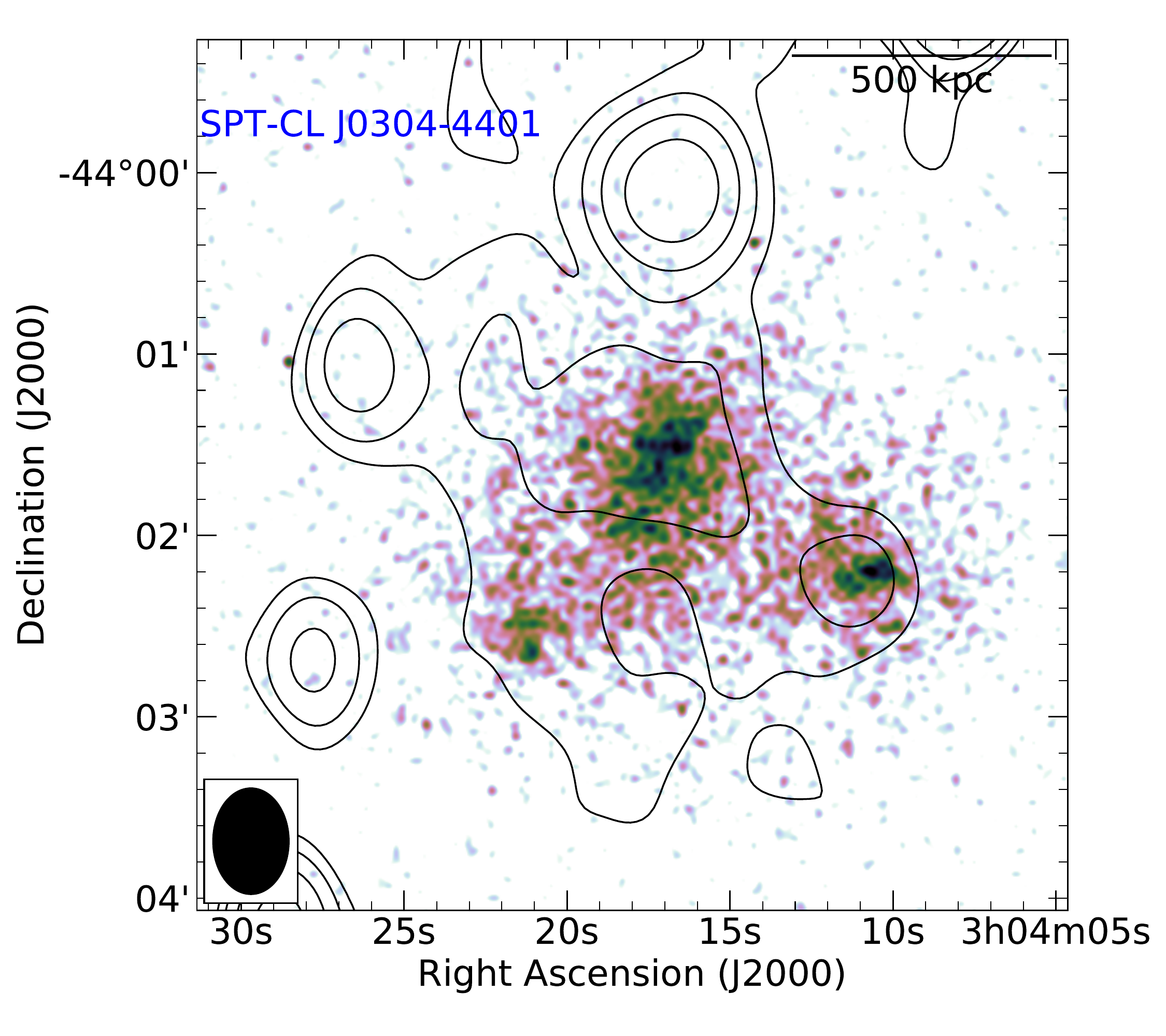} &
    \includegraphics[width=\columnwidth]{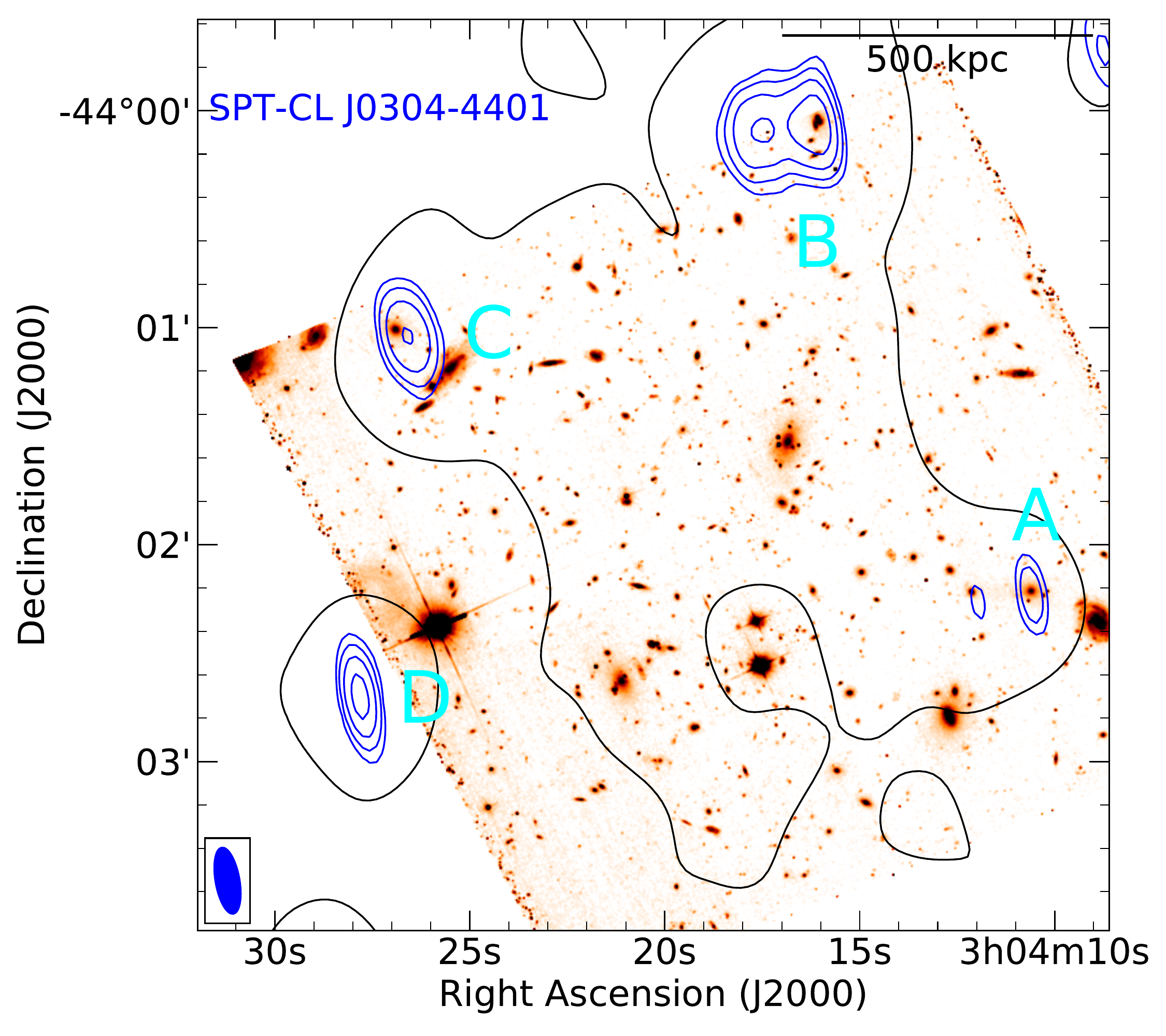}
    \end{tabular}
    \caption{\textit{Left}: \textit{Chandra} X-ray image overlaid with 325 MHz radio contours of the SPT-CL J0304-4401 cluster. The restoring beam of the radio image is $35\arcsec \times 25\arcsec$, PA $0^{\circ}$, indicated at the bottom left corner. The black contours are drawn at levels $[-1, 1, 2, 4, 8,...] \times 3\sigma_{\mathrm{rms}}$ with $\sigma_{\mathrm{rms}}$ = 180 $\mu$Jy beam$^{-1}$. Negative contours are indicated with dotted lines. 
    \textit{Right}: HST optical image is overlaid with 325 MHz high-resolution image contours excluding baselines shorter than 3k$\lambda$ (blue), and are drawn at the same levels as previously but with $\sigma_{\mathrm{rms}}$ = 80 $\mu$Jy beam$^{-1}$. The restoring beam of the radio image is $19\arcsec \times 7\arcsec$, PA $10^{\circ}$, indicated at the bottom left corner. The 3$\sigma_{\mathrm{rms}}$ contour of the \textit{left panel} is shown in black to indicate the extent of the diffuse emission. }
    \label{fig:0304-4401}
\end{figure*}

The 325 MHz images are presented in Fig. \ref{fig:0304-4401}. 
The diffuse radio emission covers almost all the cluster region visible in the X-ray image. 
The size of this diffuse emission is about $2.8\arcmin \times 3.3\arcmin$ or $1 \times 1.1$ Mpc (E-W $\times$ N-S). All the discrete point sources visible in the high-resolution radio image (blue contours in Fig. \ref{fig:0304-4401} \textit{right panel}) have optical counterparts in the HST (Hubble Space Telescope) image. In the HST image, optical counterparts are present at the peak positions of all three X-ray clumps, indicating these being sub-clusters or galaxy groups. However, only the galaxy ($z = 0.4549$; \citealt{Bayliss2016ApJS..227....3B}) corresponding to the central gas clump has spectroscopic information. The diffuse emission flux density at 325 MHz was found to be $16.65 \pm 1.85$ mJy. Considering the cluster morphology and the diffuse radio structure, we classify this extended emission as a giant radio halo.  

\subsubsection{SPT-CL J2031-4037}
The discovery of the SPT-CL J2031-4037 or RXC J2031.8-4037 cluster was first reported by \citet{Bohringer2004A&A...425..367B} in the REFLEX (ROSAT-ESO Flux Limited X-ray) Galaxy Cluster survey.
Later detections of this cluster via the Sunyaev-Zel'dovich effect were reported by \citet{Williamson2011ApJ...738..139W,Bleem2015ApJS..216...27B} and \citet{Planck2014A&A...571A..29P}.
The SPT-CL J2031-4037 is a massive ($M_{500}$ = $(9.8 \pm 1.2) \times 10^{14}$ $M_{\odot}$; \citealt{Bleem2015ApJS..216...27B}) cluster situated at the redshift $z = 0.3416$ \citep{Bohringer2004A&A...425..367B}. The \textit{Chandra} X-ray luminosity of this cluster is $L_{[0.1-2.4\ \mathrm{keV}]} = (6.8 \pm 0.3) \times 10^{44}$ erg s$^{-1}$. A previous study by \citet{Raja2020MNRAS.493L..28R} classified this as a moderately disturbed weak cool core cluster.

\begin{figure*}
	\begin{tabular}{cc}
	\includegraphics[width=\columnwidth]{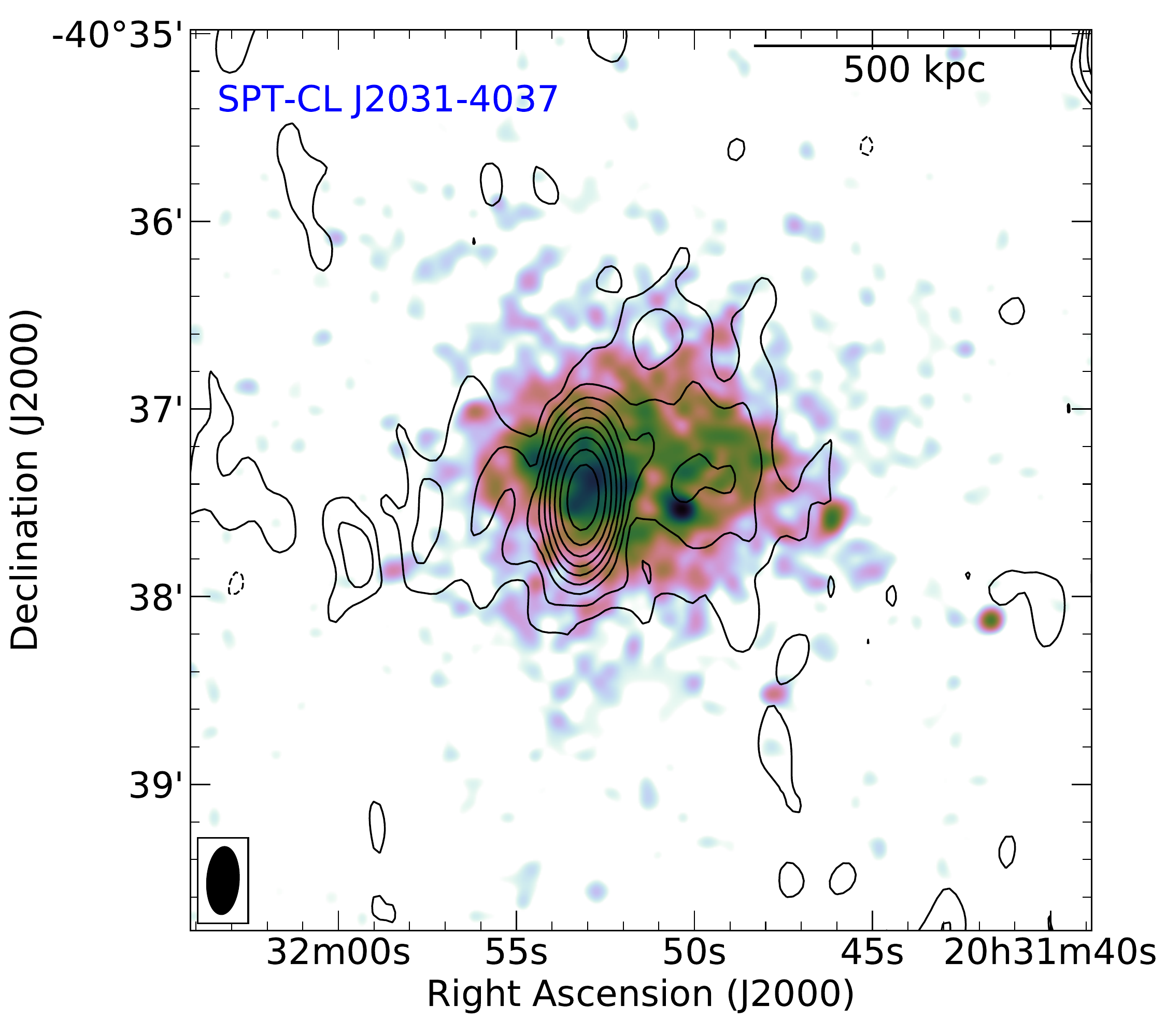} &
	\includegraphics[width=\columnwidth]{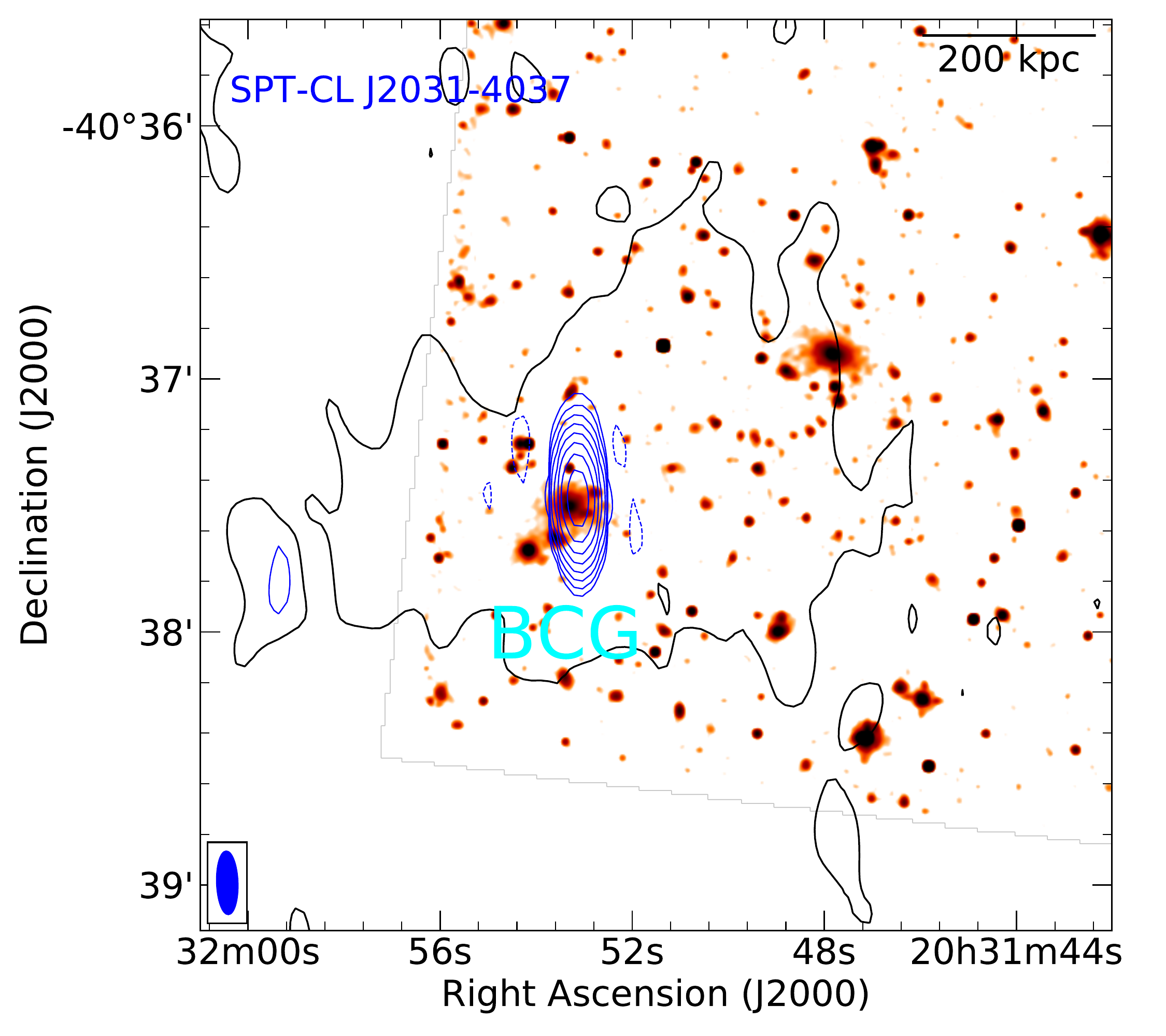}
	\end{tabular}
	\caption{\textit{Left}: \textit{Chandra} X-ray image overlaid with 325 MHz radio contours of the SPT-CL J2031-4037 cluster. The restoring beam of the radio image is $21\arcsec \times 10\arcsec$, PA $-4^{\circ}$, indicated at the bottom left corner. The black contours are drawn at levels $[-1, 1, 2, 4, 8,...] \times 3\sigma_{\mathrm{rms}}$ with $\sigma_{\mathrm{rms}}$ = 60 $\mu$Jy beam$^{-1}$. Negative contours are indicated with dotted lines. \textit{Right}: HST optical image is overlaid with the 325 MHz high resolution image contours excluding baselines shorter than $5$k$\lambda$ (blue) and are drawn at the same levels as previously but with $\sigma_{\mathrm{rms}}$ = 120 $\mu$Jy beam$^{-1}$. The restoring beam of the radio image is $15\arcsec \times 5\arcsec$, PA $2^{\circ}$, indicated at the bottom left corner. The 3$\sigma_{\mathrm{rms}}$ contour of the \textit{left panel} is shown in black to show the extent of the diffuse emission.}
    \label{fig:2031-4037}
\end{figure*}

A multi-frequency study of the diffuse emission in this cluster was carried out by \citet{Raja2020MNRAS.493L..28R}, and only a brief description of this source is presented here.
The 325 MHz images are presented in Fig. \ref{fig:2031-4037}. The diffuse radio emission is present throughout most of the cluster region visible in the X-ray image. It has a similar east-west extension as the ICM seen in the X-ray image, which is possibly the merger axis. The size of the diffuse emission is about $2.7\arcmin \times 2.1\arcmin$ or $0.8 \times 0.6$ Mpc (E-W $\times$ N-S). A bright radio galaxy (BCG) is present at the position of the X-ray peak, which also has optical identification in the HST optical image. The BCG was also detected in the TGSS \citep{Intema2017A&A...598A..78I} and SUMSS \citep{Bock1999AJ....117.1578B,Mauch2003MNRAS.342.1117M}. The GLEAM survey \citep{Wayth2015PASA...32...25W,Hurley-Walker2017MNRAS.464.1146H} also detected a blob at this position, which encompasses the entire cluster region.
Considering the morphology, and the diffuse radio structure, it was classified as an intermediate radio halo with the flux density of $16.93 \pm 1.76$ mJy at 325 MHz \citet{Raja2020MNRAS.493L..28R}.

\subsubsection{SPT-CL J2248-4431}
The SPT-CL J2248-4431 cluster or otherwise known as ACO S 1063 was first discovered by \citet{Abell1989ApJS...70....1A}. Subsequent detections of this cluster were reported in REFLEX survey by \citet{Bohringer2004A&A...425..367B}, in SPT-SZ survey by \citet{Williamson2011ApJ...738..139W}, in \textit{Planck} SZ survey by \citet{Planck2014A&A...571A..29P} and more.
This is a massive ($M_{500}$ = $(18.0 \pm 2.2) \times 10^{14}$ $M_{\odot}$; \citealt{Bleem2015ApJS..216...27B}) cluster situated at the redshift $z = 0.351$ \citep{Bleem2015ApJS..216...27B}. Optical and X-ray study performed by \citet{Gomez2012AJ....144...79G} reported this cluster to be bullet like merger with the merger axis being north-east to south-west, and the merging plane is close the sky plane.
It is an extremely X-ray luminous cluster with $L_{[0.1-2.4\ \mathrm{keV}]} = (25.1 \pm 0.3) \times 10^{44}$ erg s$^{-1}$, and the second most luminous cluster in the REFLEX survey.
A detailed X-ray study done by Rahaman et al. (submitted to ApJ), with X-ray temperature map and morphology parameters, classified this as a moderately disturbed weak cool core (WCC) cluster which is in transition into a non-cool core (NCC) cluster.

\begin{figure*}
\begin{tabular}{cc}
\includegraphics[width=\columnwidth]{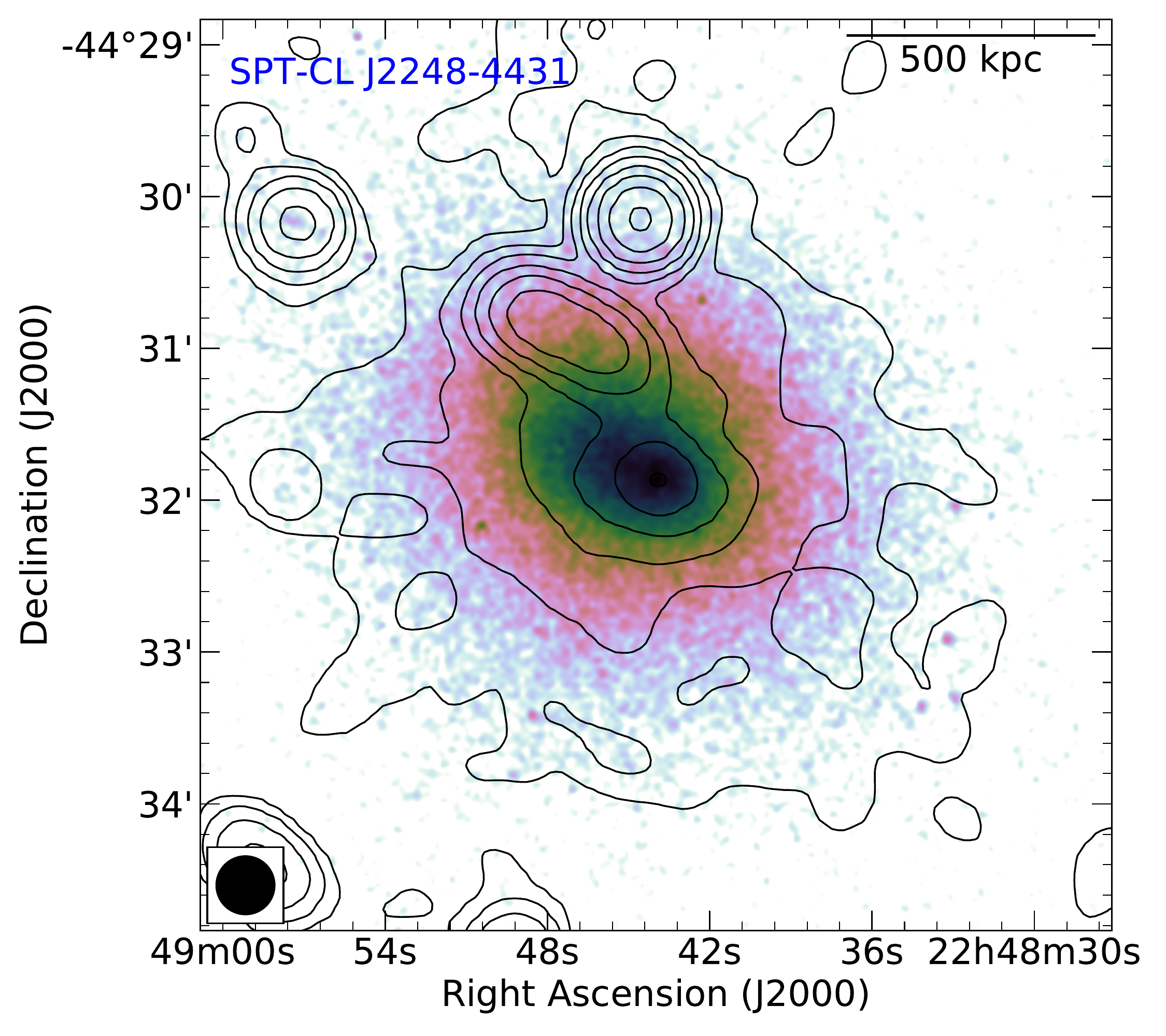} &
\includegraphics[width=\columnwidth]{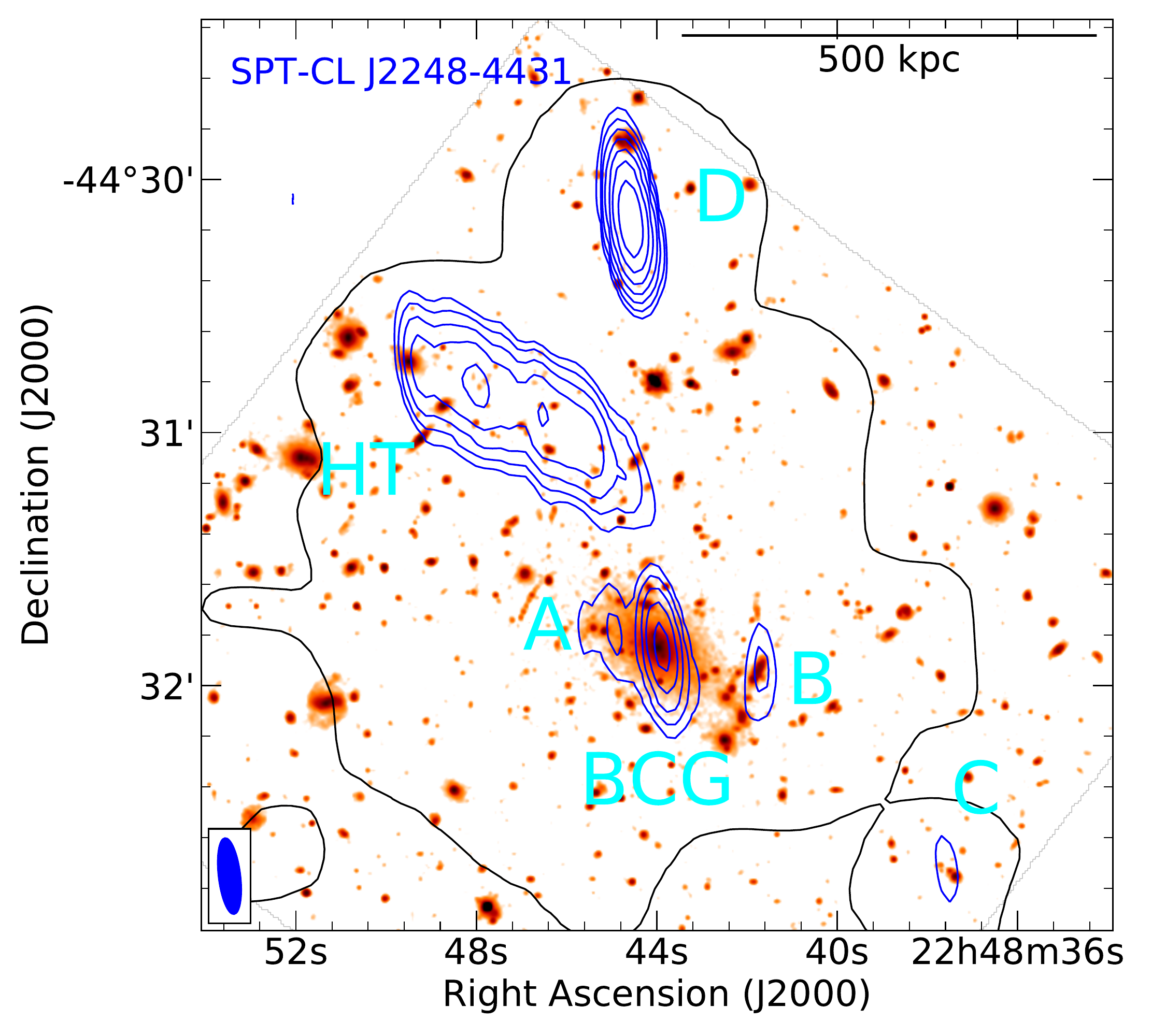} 
\end{tabular}
\caption{\textit{Left}: \textit{Chandra} X-ray image overlaid with 325 MHz radio contours of the SPT-CL J2248-4431 cluster. The restoring beam of the radio image is $23\arcsec \times 23\arcsec$, PA $0^{\circ}$, indicated at the bottom left corner. The black contours are drawn at levels $[-1, 1, 2, 4, 8,...] \times 3\sigma_{\mathrm{rms}}$ with $\sigma_{\mathrm{rms}}$ = 100 $\mu$Jy beam$^{-1}$. Negative contours are indicated with dotted lines. 
\textit{Right}: HST optical image is overlaid with 325 MHz high resolution image contours excluding baselines shorter than 1k$\lambda$ (blue) and are drawn at the same levels as previously but with $\sigma_{\mathrm{rms}}$ = 100 $\mu$Jy beam$^{-1}$. The restoring beam of the radio image is $18\arcsec \times 5\arcsec$, PA $7^{\circ}$, indicated at the bottom left corner. The 6$\sigma_{\mathrm{rms}}$ contour of the \textit{left panel} is shown in black to show the extent of the diffuse emission.}
\label{fig:2248-4431}
\end{figure*}

Recently, multi-frequency observations carried out by \citet{Xie2020arXiv200104725X} reported the detection of diffuse radio emission in this cluster at 325 MHz along with 1.5 and 3.0 GHz and classified it as a giant radio halo. Here, we have presented only a brief description of the observed diffuse emission at 325 MHz, and a more detailed analysis is presented in Rahaman et al. (submitted to ApJ).
In Fig. \ref{fig:2248-4431}, we present the \textit{Chandra} X-ray, 325 MHz radio, and HST optical image of the SPT-CL J2248-4431 cluster. It can be seen that the diffuse radio emission covers the entire cluster region visible in X-ray. The extent of this extended emission was found to be $\sim$1.2 Mpc. The HST image shows the optical counterparts of all the radio galaxies visible in the 325 MHz high-resolution image (blue contours) in Fig. \ref{fig:2248-4431} \textit{right panel}. The flux density of the diffuse radio emission excluding the point sources and the head-tail radio galaxy was found to be $62.0 \pm 6.28$ mJy at 325 MHz. This halo flux density is much higher than what was reported by \citet{Xie2020arXiv200104725X}, and the reliability of our flux density estimation is described in detail in Rahaman et al. (submitted to ApJ). We tried a combination of different imaging parameters as well as point source modelling corresponding to different inner uv-cut. In all cases, the diffuse flux densities were found consistent with the one reported here. Radio emission from this cluster was detected in other surveys as well. 
In the TGSS \citep{Intema2017A&A...598A..78I} and SUMSS \citep{Bock1999AJ....117.1578B,Mauch2003MNRAS.342.1117M} survey, a blob emission was detected consisting of the cluster radio galaxies. In the GLEAM \citep{Wayth2015PASA...32...25W,Hurley-Walker2017MNRAS.464.1146H} survey, a single blob-like structure was detected covering the entire cluster region.

\subsection{Diffuse emission candidate} \label{subsec:partial_detect}
\subsubsection{SPT-CL J0348-4515}
\begin{figure*}
    \begin{tabular}{cc}
    \includegraphics[width=\columnwidth]{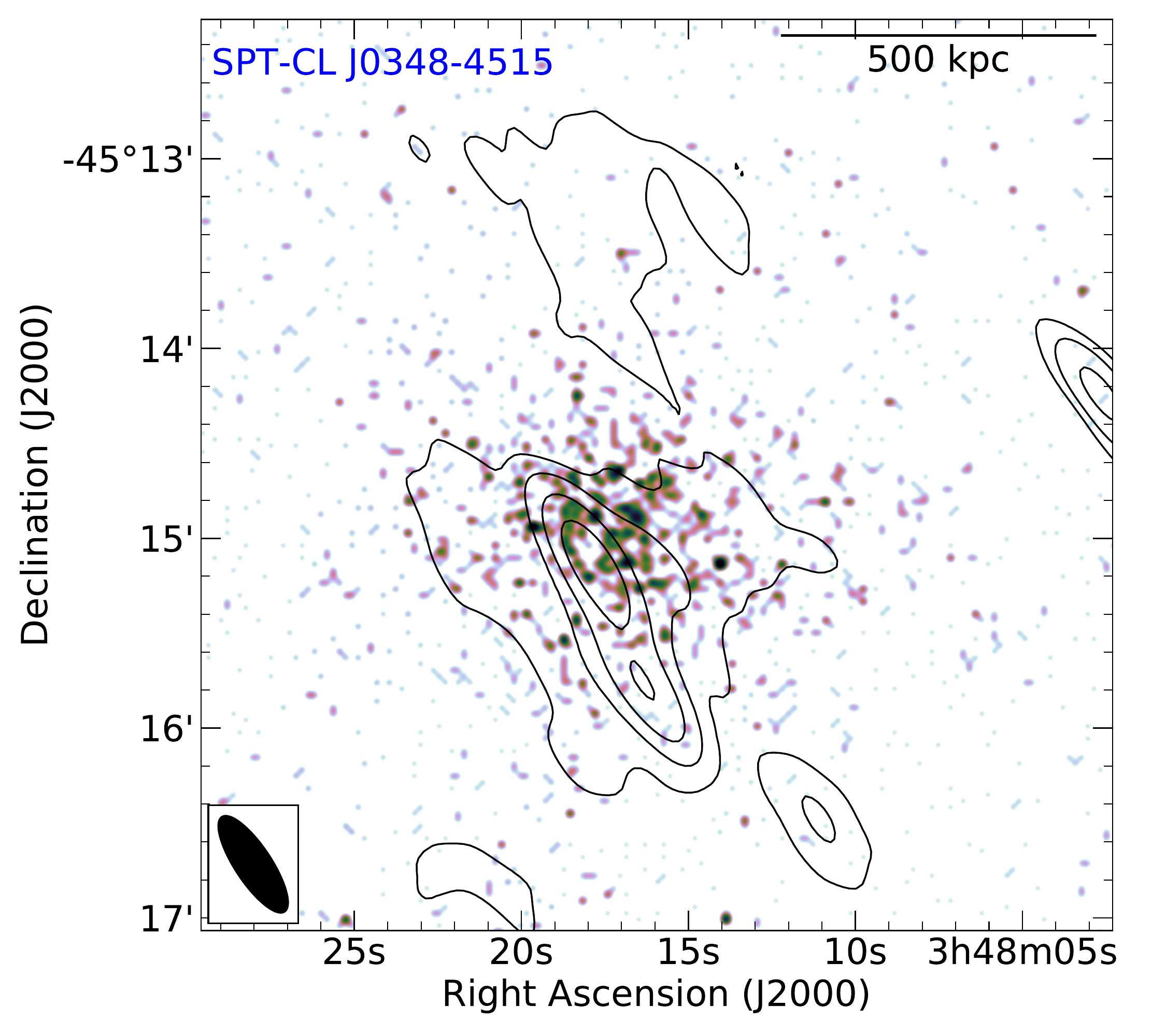} &
    \includegraphics[width=\columnwidth]{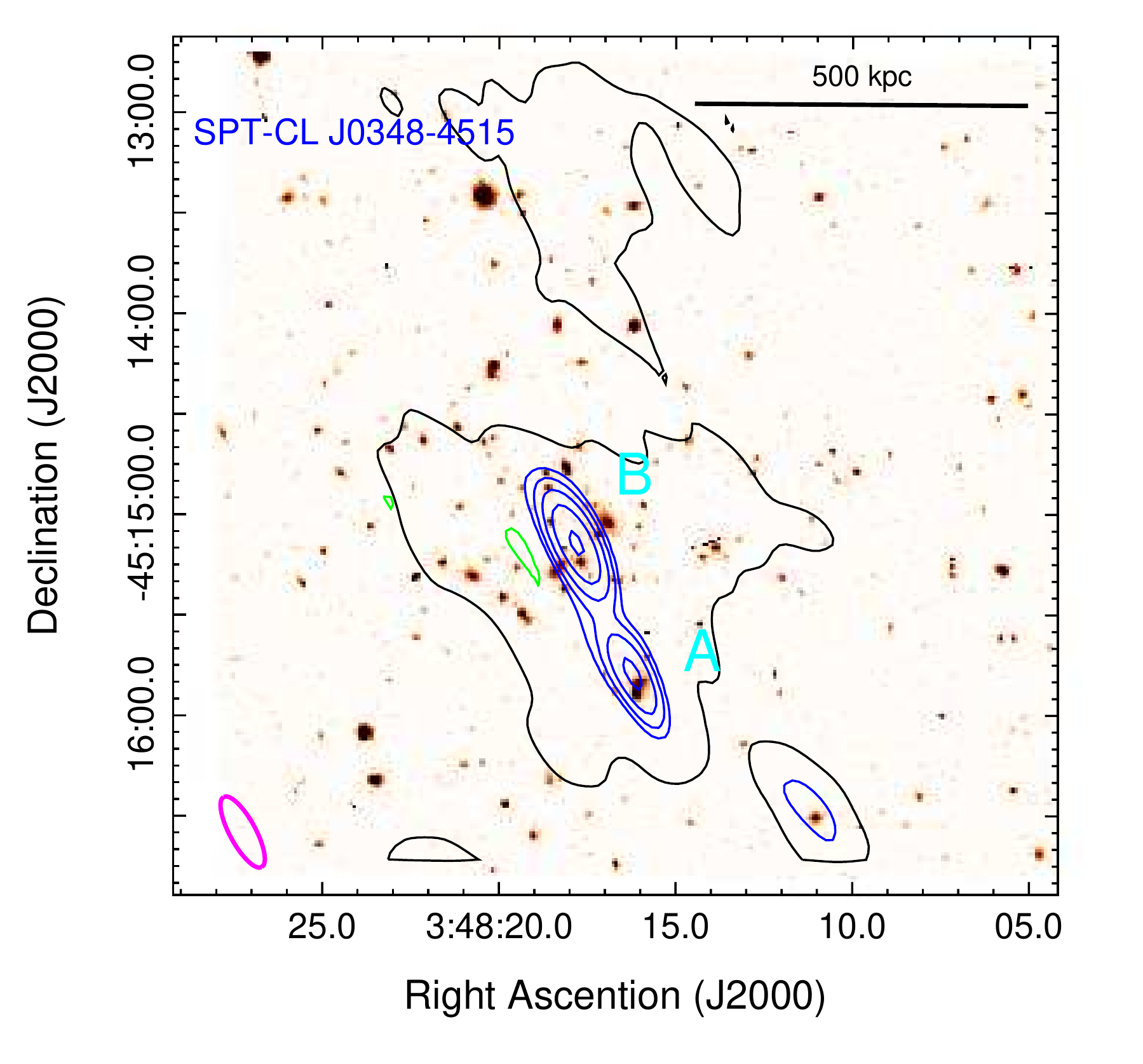}
    \end{tabular}
    \caption{\textit{Left}: \textit{Chandra} X-ray image overlaid with 325 MHz radio contours of the SPT-CL J0348-4515 cluster. The restoring beam of the radio image is $36\arcsec \times 11.5\arcsec$, PA $33.5^{\circ}$, indicated at the bottom left corner. The black contours are drawn at levels $[-1, 1, 2, 4, 8,...] \times 3\sigma_{\mathrm{rms}}$ with $\sigma_{\mathrm{rms}}$ = 150 $\mu$Jy beam$^{-1}$. Negative contours are indicated with dotted lines. 
    \textit{Right}: DECam optical image is overlaid with 325 MHz high-resolution image contours excluding baselines shorter than 3k$\lambda$ (blue) and are drawn at the same levels as previously but with $\sigma_{\mathrm{rms}}$ = 100 $\mu$Jy beam$^{-1}$. Negative contours are indicated in green. The restoring beam of the radio image is $24\arcsec \times 8\arcsec$, PA $29^{\circ}$, indicated at the bottom left corner. The 3$\sigma_{\mathrm{rms}}$ contour of the \textit{left panel} is shown in black to show the extent of the diffuse emission.}
    \label{fig:0348-4515}
\end{figure*}

The discovery of the SPT-CL J0348-4515 or ClG 0346-454 cluster was first reported by \citet{West1981A&AS...44..329W} with the SRC Schmidt plate.
This is a massive ($M_{500}$ = $(6.2 \pm 1.0) \times 10^{14}$ $M_{\odot}$; \citealt{Bleem2015ApJS..216...27B}) cluster situated at the redshift $z = 0.358$ \citep{Bleem2015ApJS..216...27B}. The \textit{Chandra} X-ray image (Fig. \ref{fig:0348-4515} \textit{left panel}) shows an irregular ICM distribution and the X-ray luminosity of this cluster was found to be $L_{[0.1-2.4\ \mathrm{keV}]} = (3.2 \pm 0.3) \times 10^{44}$ erg s$^{-1}$. The central temperature and morphology parameters suggest that it is a disturbed weak cool core cluster (Table \ref{tab:sample}).

The 325 MHz images are presented in Fig. \ref{fig:0348-4515}. The diffuse radio emission is present throughout most of the cluster region visible in the X-ray image. The size of this diffuse emission is about $2.3\arcmin \times 1.9\arcmin$ or $0.7 \times 0.6$ Mpc (E-W $\times$ N-S). 
Both radio galaxies in the cluster region visible in the high-resolution image contours have optical counterparts in the DECam image. 
After subtracting the contributions from the radio galaxies, the diffuse flux density comes out to be $12.04 \pm 1.38$ mJy at 325 MHz. We classify this emission to be a candidate halo.
In addition, another diffuse radio emission of $\sim0.5$ Mpc scale is present in the north of the cluster, which does not have any corresponding optical counterpart. The flux density of this diffuse source was found to be $4.43 \pm 0.62$ mJy. The position of this diffuse source probably excludes it being a part of the halo emission; however, the possibility of it being revived fossil plasma can not be ruled out. Further deeper observations are needed to confirm these possibilities. 

\begin{figure*}
    \begin{tabular}{cc}
    \includegraphics[width=0.97\columnwidth]{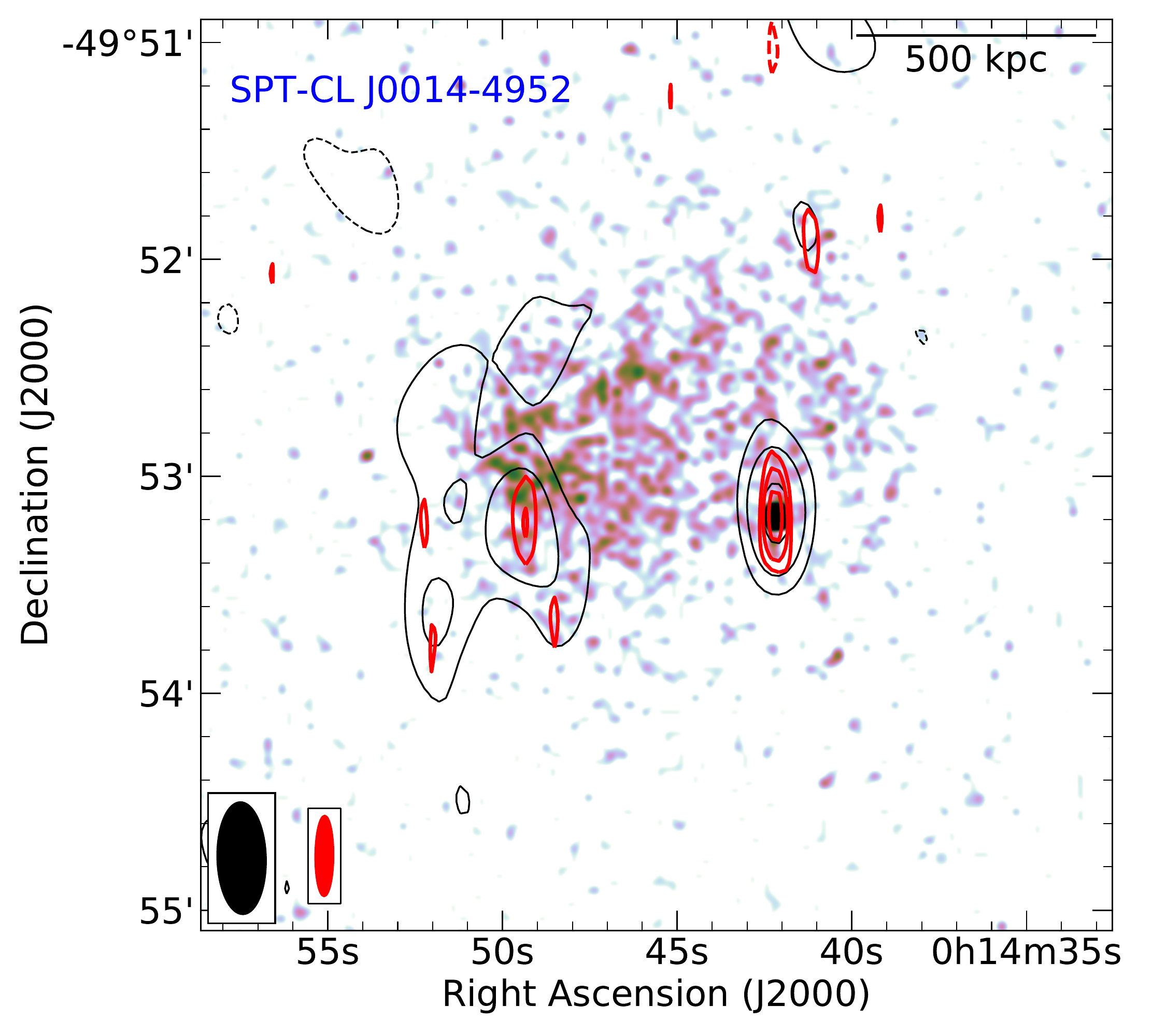} &
    \includegraphics[width=0.97\columnwidth]{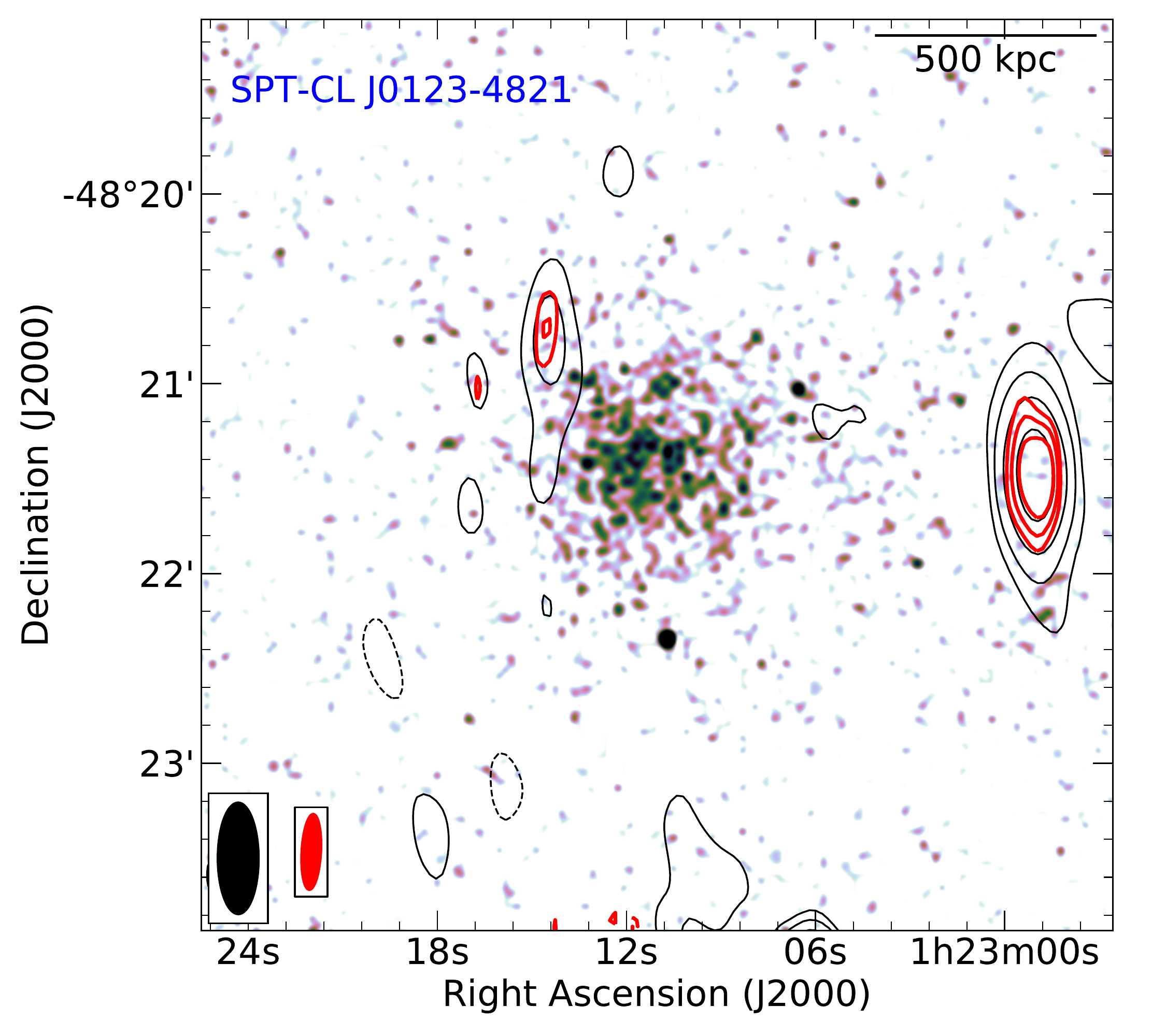} \\
    \includegraphics[width=0.97\columnwidth]{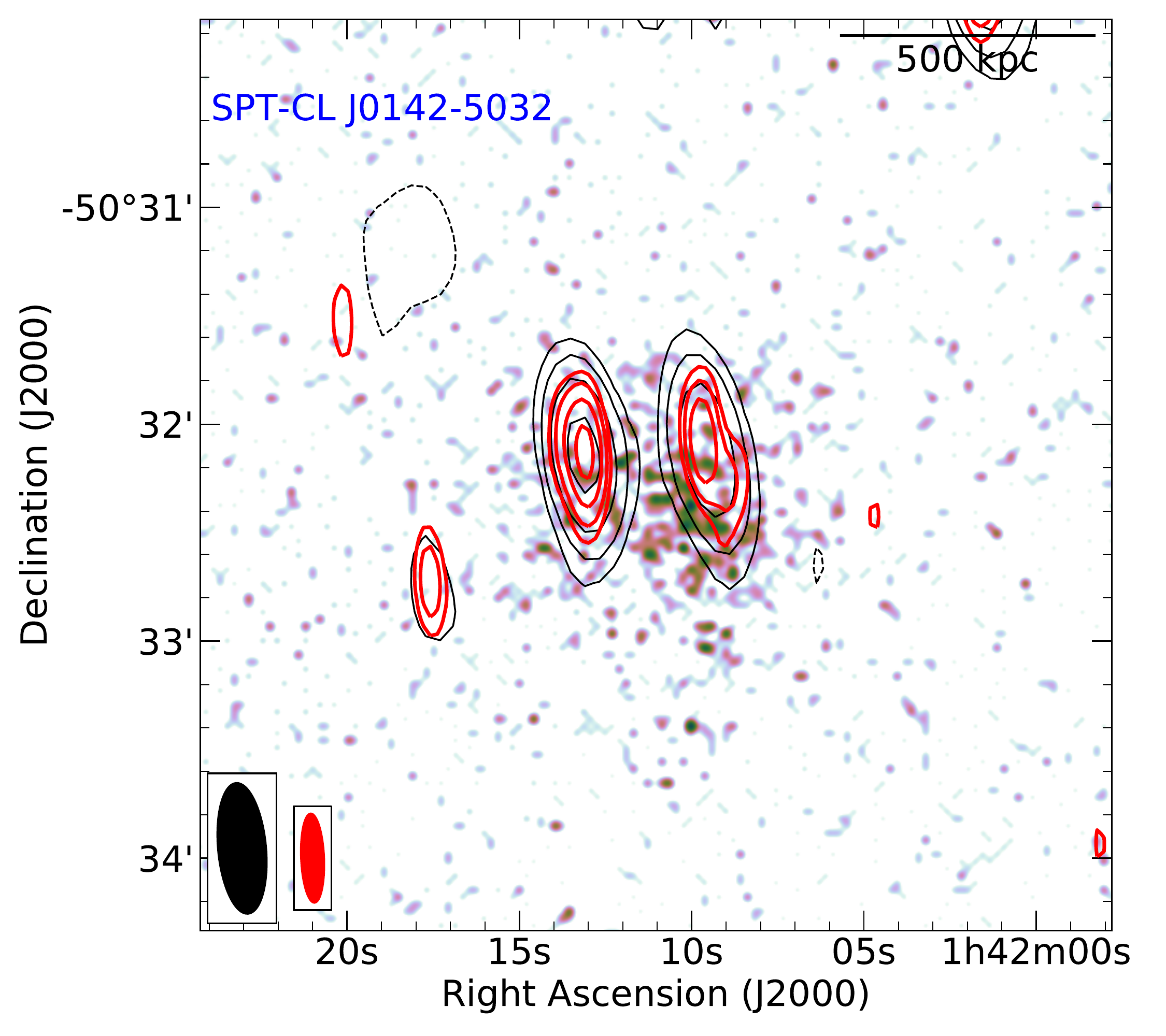} &
    \includegraphics[width=0.97\columnwidth]{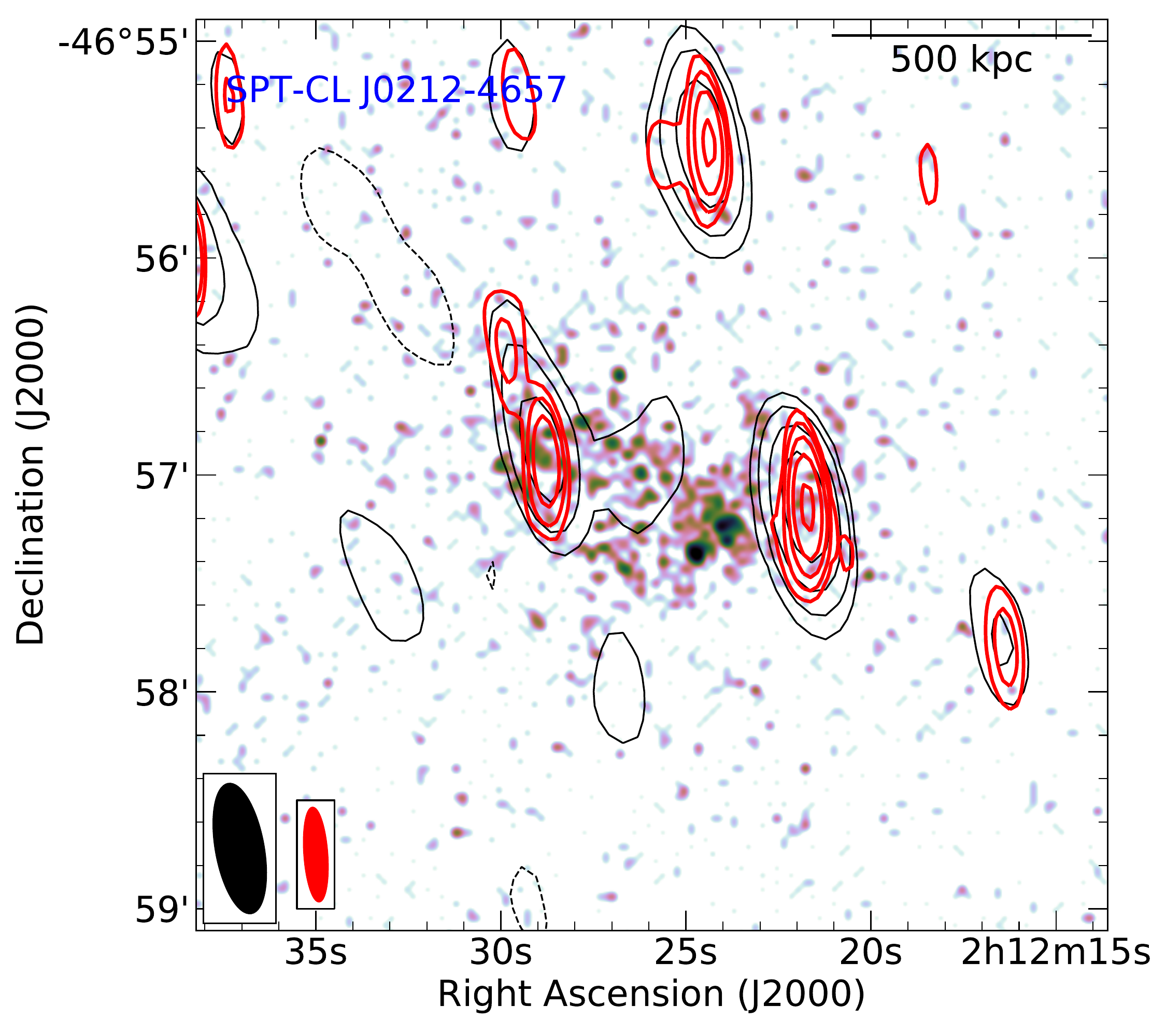} \\
    \includegraphics[width=0.97\columnwidth]{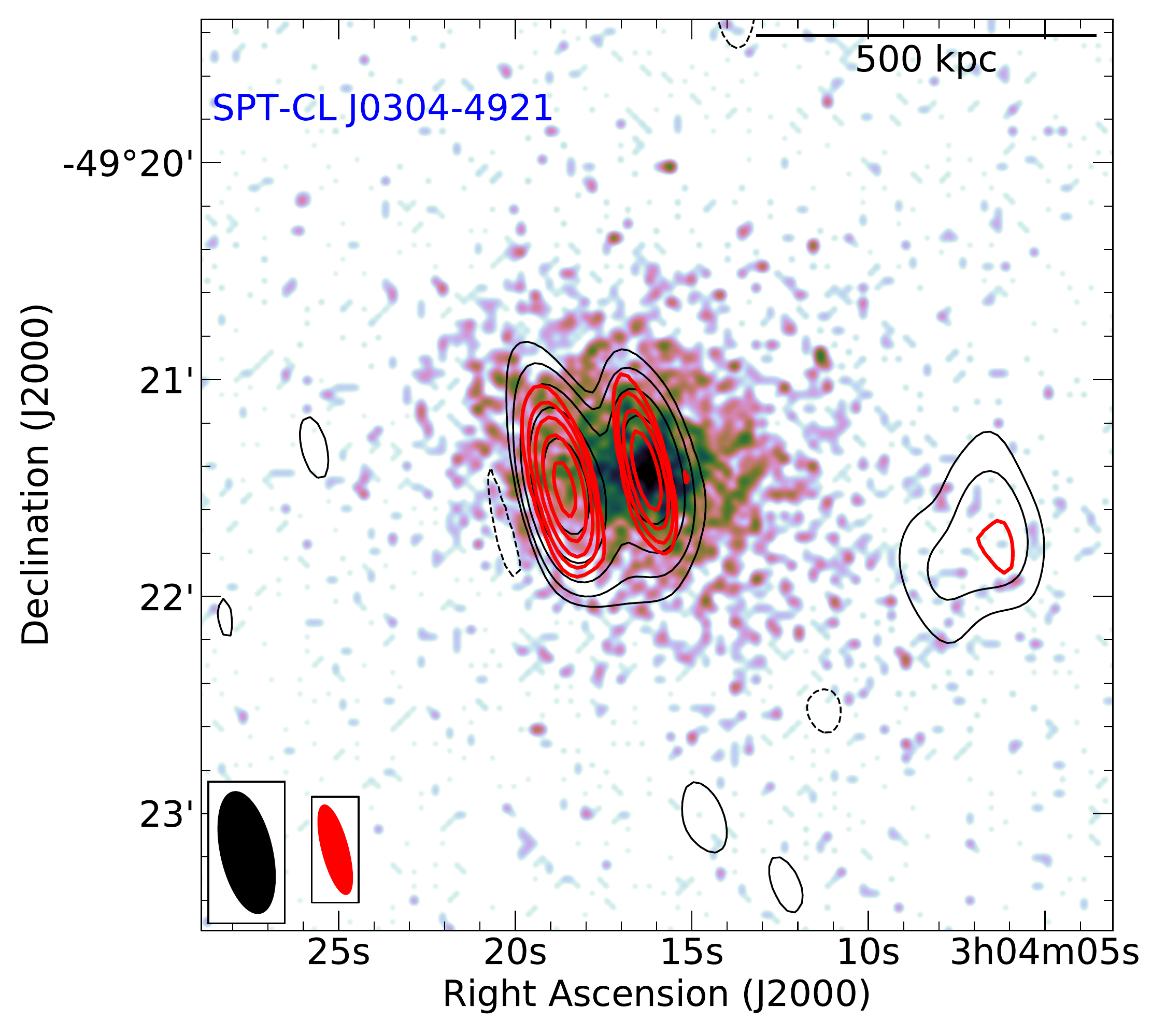} &
    \includegraphics[width=0.97\columnwidth]{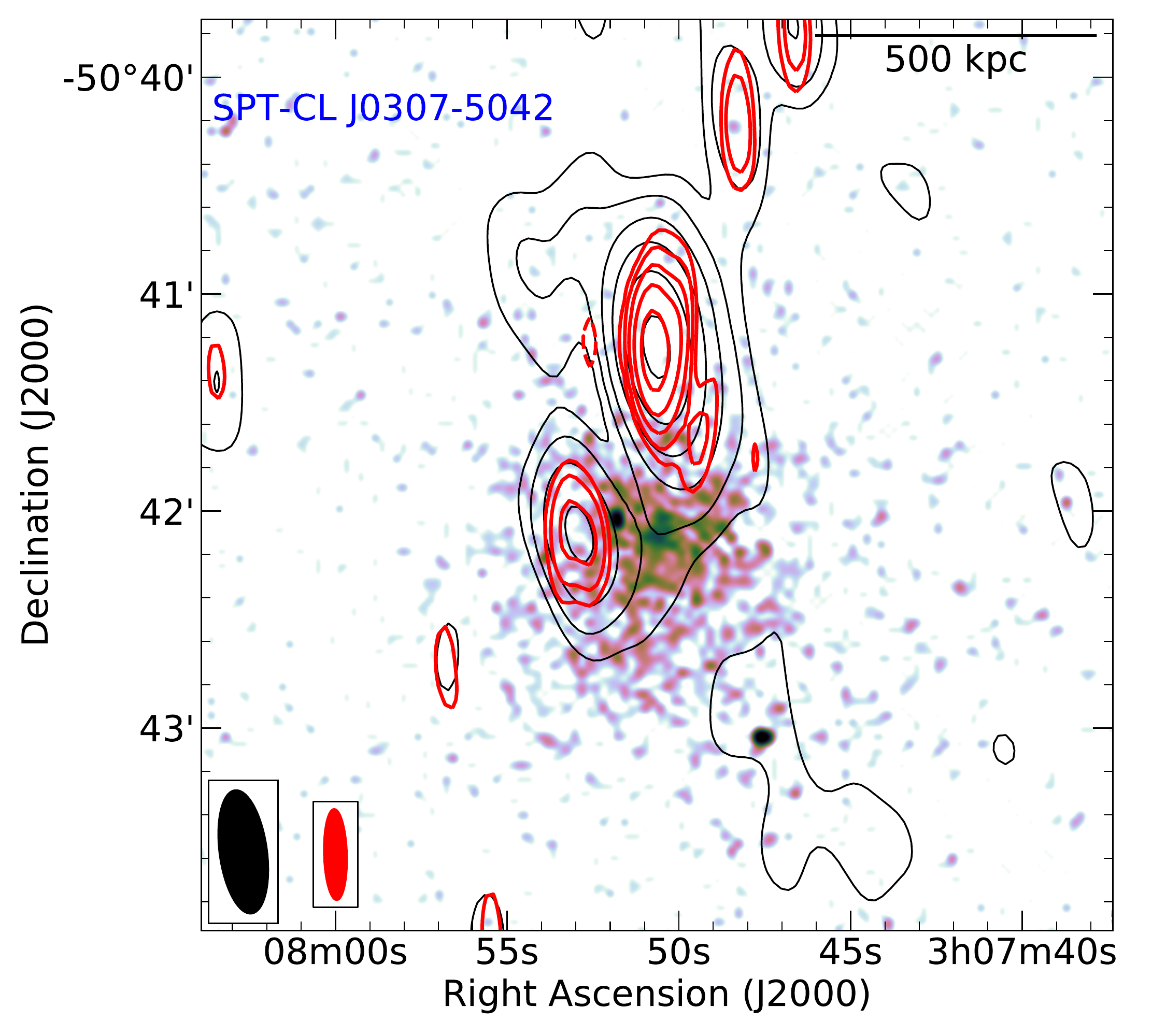}
    \end{tabular}
    \caption{\textit{Chandra} X-ray images overlaid with 325 MHz radio contours of the SPT-CL J0014-4952, SPT-CL J0123-4821, SPT-CL J0142-5032, SPT-CL J0212-4657, SPT-CL J0304-4921, and SPT-CL J0307-5042. The contours are drawn at levels $[-1, 1, 2, 4, 8,...] \times 3\sigma_{\mathrm{rms}}$. Negative contours are indicated with dotted lines. The restoring beam of the low-resolution and high-resolution images are indicated in the bottom left corner with black and red ellipses, respectively. See Table \ref{tab:Image_info} for $\sigma_{\mathrm{rms}}$ and restoring beams.}
    \label{fig:ULs1}
\end{figure*}

\begin{figure*}
    \begin{tabular}{cc}
    \includegraphics[width=\columnwidth]{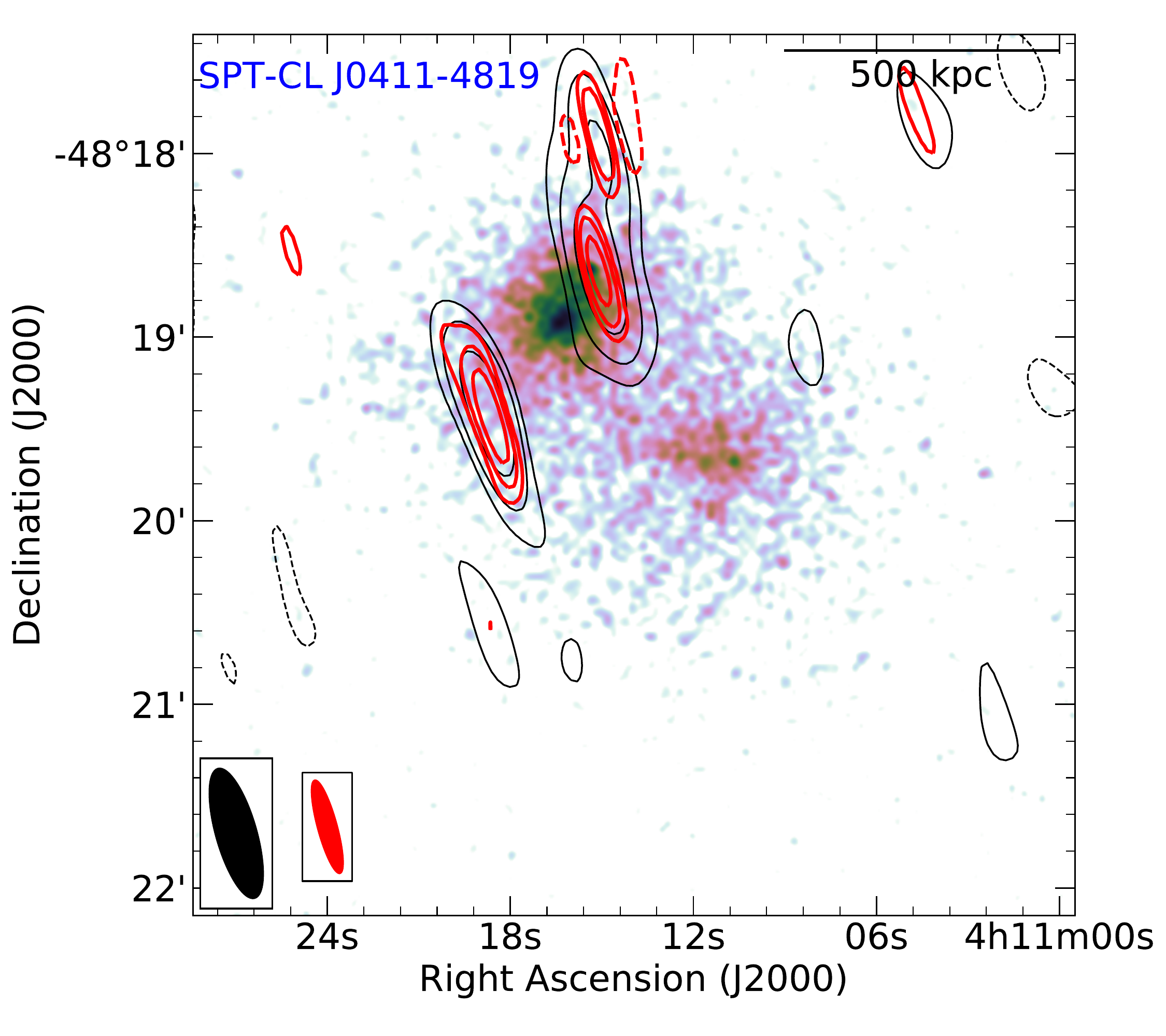} &
    \includegraphics[width=\columnwidth]{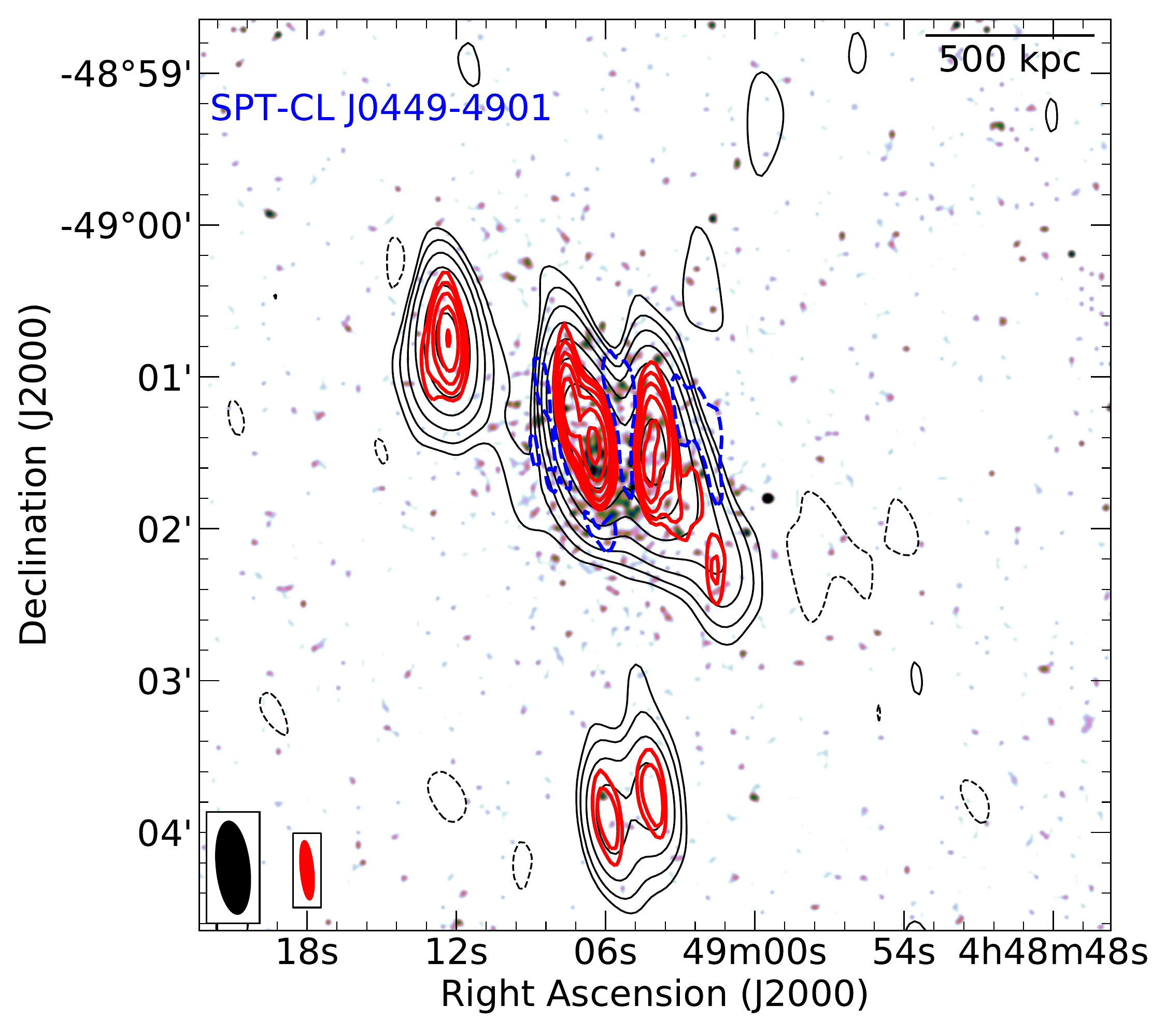} \\
    \includegraphics[width=\columnwidth]{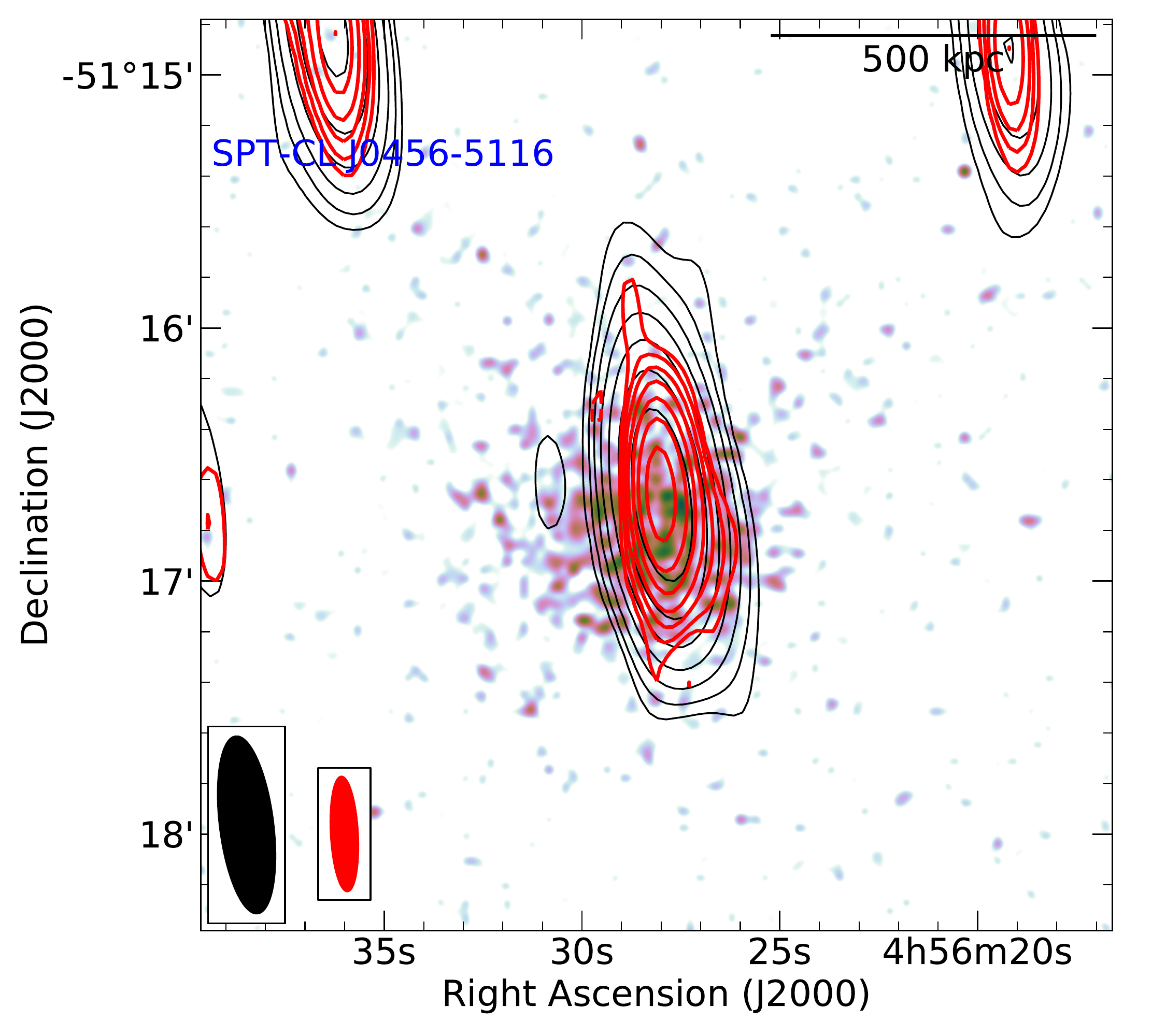} &
    \includegraphics[width=\columnwidth]{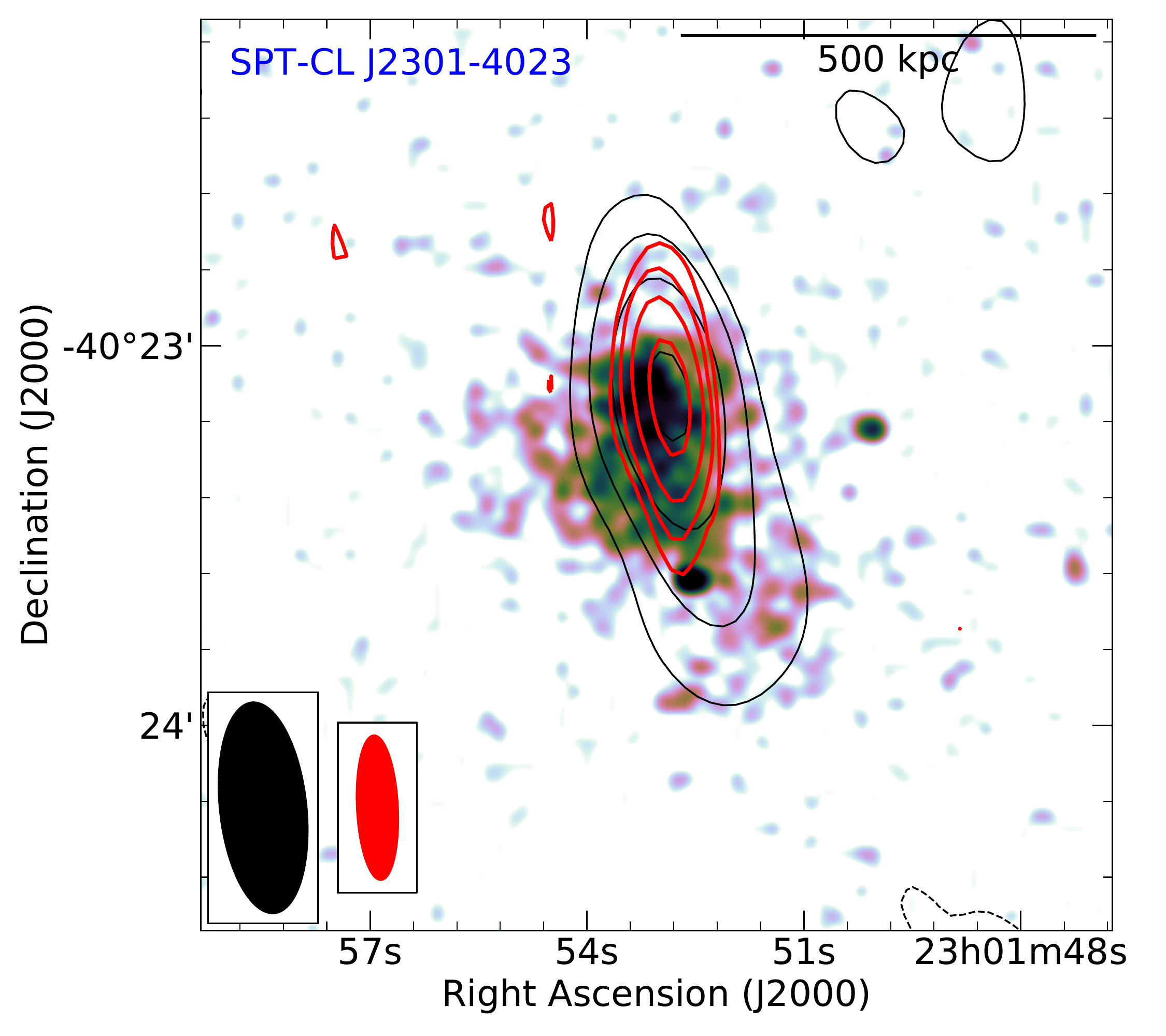} 
    \end{tabular}
    \caption{\textit{Chandra} X-ray image overlaid with 325 MHz radio contours of the SPT-CL J0411-4819, SPT-CL J0449-4901, SPT-CL J0456-5116, and SPT-CL J2301-4023. The contours are drawn at levels $[-1, 1, 2, 4, 8,...] \times 3\sigma_{\mathrm{rms}}$. Negative contours are indicated with dotted lines. The restoring beam of the low-resolution and high-resolution images are indicated in the bottom left corner with black and red ellipses, respectively. See Table \ref{tab:Image_info} for $\sigma_{\mathrm{rms}}$ and restoring beams.}
    \label{fig:ULs2}
\end{figure*}


\begin{table*}
    \caption{Flux densities of the discrete and diffuse radio sources of the cluster sample}
    \label{tab:flux}
    
    \begin{tabular}{lccccc}
    \hline
    Name & Source & $\mathrm{RA_{J2000}}$ $\mathrm{DEC_{J2000}}$ & $S_{\nu}$ & $P_\mathrm{{1.4\ GHz}}$ & Size \\
    \noalign{\smallskip}
     & & hh mm ss\  $^\circ\ \arcmin\ \arcsec$ & (mJy) & ($10^{24}$ W Hz$^{-1}$) & (Mpc) \\
    \hline
    SPT-CL J0013-4906 & halo & & $11.01 \pm 1.19$ & $1.07 \pm 0.12$ & $0.89$ \\
    \hline
    SPT-CL J0014-4952 & halo (UL) & & $<5$ & $<3.67$ &  \\ 
    \hline
    SPT-CL J0123-4821 & halo (UL) & & $<4$ & $<1.77$ &  \\
    \hline
    SPT-CL J0142-5032 & halo (UL) & & $<10$ & $<2.82$ & \\
    \hline
    SPT-CL J0212-4657 & halo (UL) & & Bad rms &  &  \\
    \hline
    SPT-CL J0304-4401 & A & 03 04 10.76 -44 02 13.43 & $0.73 \pm 0.07$ & & \\
     & B & 03 04 16.15 -44 00 04.40 & $9.35 \pm 0.32$ & & \\
     & C & 03 04 26.55 -44 01 03.93 & $3.25 \pm 0.13$ & & \\
     & D & 03 04 27.83 -44 02 41.65 & $2.72 \pm 0.09$ & & \\
     & halo & & $16.65 \pm 1.85$ & $2.17 \pm 0.24$ & 1.07 \\
    \hline
    SPT-CL J0304-4921 & halo (UL) & & $<4$ & $<0.53$ & \\
    \hline
    SPT-CL J0307-5042 & halo (UL) & & $<5$ & $<2.06$ &  \\
    \hline
    SPT-CL J0348-4515 & A & 03 48 16.30 -45 15 47.56 & $3.01 \pm 0.16$ & & \\
     & B & 03 48 17.85 -45 15 08.96 & $6.26 \pm 0.17$ & & \\
     & halo (C) & & $12.04 \pm 1.38$ & $0.84 \pm 0.1$ & 0.64 \\
    \hline
    SPT-CL J0411-4819 & halo (UL) & & Bad rms &  &  \\
    \hline
    SPT-CL J0449-4901 & halo (UL) & & $<4$ & $<4.20$ & \\
    \hline
    SPT-CL J0456-5116 & halo (UL) & & $<10$ & $<3.05$ & \\
    \hline
    SPT-CL J2031-4037 & BCG & 20 31 51.49  -40 37 14.02 & $89.6 \pm 0.53$ & & \\
     & halo & & $16.93 \pm 1.76$ & $0.77 \pm 0.08^{\dagger}$ & 0.69 \\
    \hline
    SPT-CL J2248-4431 & A & 22 48 44.95 -44 31 47.86 & $0.33 \pm 0.09$ & & \\
     & BCG & 22 48 43.86 -44 31 51.33 & $6.61 \pm 0.11$ & & \\
     & B & 22 48 41.70 -44 31 57.12 & $0.54 \pm 0.1$ & & \\
     & C & 22 48 37.58 -44 32 43.40 & $0.58 \pm 0.12$ & & \\
     & D & 22 48 44.62 -44 30 08.90 & $42.49 \pm 0.18$ & & \\
     & RG & 22 48 49.06 -44 30 44.19 & $36.48 \pm 0.4$ & & \\
     & halo & & $62.0 \pm 6.28$ & $4.22 \pm 0.43$ & 1.2 \\
    \hline
    SPT-CL J2301-4023 & halo (UL) & & $<7$ & $<6.01$ &  \\
    \hline
    \end{tabular}
\\\flushleft{\textit{Note.} The columns are 1. Cluster name, 2. Radio source, 3. Right Ascension, Declination, 4. Flux density at 325 MHz, 5. Radio power at 1.4 GHz assuming spectral index $\alpha = -1.3$, 6. Size of the diffuse emission ($\mathrm{\sqrt{E-W \times N-S}}$). $^{\dagger}$The 1.4 GHz radio power of the SPT-CL J2031-4037 halo is taken from \citet{Raja2020MNRAS.493L..28R} with the observed spectral index of $\alpha=-1.35$. Halo (C) and halo (UL) represents candidate halo and the halo upper limit, respectively.}
\end{table*}

\subsection{Non-detections and upper limits} \label{subsec:non-detect}

The clusters in the sample that do not show any presence of diffuse radio emission are classified here as non-detections. All these clusters have only one or two bright radio galaxy present within the cluster extent. 

For the determination of the radio halo upper limit, we have followed a similar approach as described in \citet{Paul2019MNRAS.489..446P}, which in turn based on the approach described in \citet{Bonafede2017MNRAS.470.3465B}. 
\begin{enumerate}
    \item First, we derive the expected radio power corresponding to the cluster mass using the correlation given in \citet{Cassano2013ApJ...777..141C} i.e.,
    \begin{equation}\label{eq:P-M}
    \log\bigg(\frac{P_{1.4}}{10^{24.5} \; \rm{W \; Hz^{-1}}}\bigg) = B \; \log \bigg( \frac{M_{500}}{10^{14.9} \: \rm{M_{\odot}} }\bigg) + A
    \end{equation}
    
    where $A=0.125 \pm 0.076$ and $B=3.77\pm0.57$ are BCES-bisector fitting parameters corresponding to `Radio Halo' only data.
    
    \item Next, we calculated the expected radio halo size from the above-derived radio power using the relation presented in \citet{Cassano2007MNRAS.378.1565C} i.e.,
    \begin{eqnarray}\label{eq:P-R}
    \log\bigg(\frac{P_{1.4}}{5 \times 10^{24} \:  h^{-2}_{70}\: \rm{W \: Hz^{-1}}}\bigg)    {}&=\; (4.18 \pm 0.68) \: \log \bigg( \frac{R_{H}}{500 h^{-1}_{70} \: \rm{kpc} }\bigg)     \nonumber \\  
    & - (0.26	\pm 0.07)
    \end{eqnarray}
    
    where $R_H$ is the radio halo raddi.
    
    \item Then, we injected a fake radio halo with the expected radio power within the expected radio halo size, which is calculated above, in the calibrated UV-data at a position near the cluster and free from other radio emission, taking into account the w-projection effect.
    
    \item We varied the injected fake radio halo flux density $S^{inj}_{R_H}$ keeping the halo size constant. Now, if the recovered halo has $D^{meas}_{2\sigma} > R_{H}$ and $S^{meas}_{2\sigma} \gg 30\% S^{inj}_{R_{H}}$, where $D^{meas}_{2\sigma}$ and $S^{meas}_{2\sigma}$ are measured fake halo size and flux density, the injected flux density is decreased. Similarly, if the halo size $D^{meas}_{2\sigma} < R_{H}$, the injected flux density is increased. This process is repeated until we reach the condition where $D^{meas}_{2\sigma} \gtrsim R_{H}$. The injected flux density at this point is considered the upper limit to the radio halo.
\end{enumerate}


A brief description of the individual clusters along with their halo upper limits, is presented below.

\subsubsection{SPT-CL J0014-4952}
The SPT-CL J0014-4952 \citep{McDonald2013ApJ...774...23M} is a less massive ($M_{500}$ = $(5.3 \pm 0.9) \times 10^{14}$ $M_{\odot}$; \citealt{Bleem2015ApJS..216...27B}) cluster situated at the redshift $z = 0.752$ \citep{Bleem2015ApJS..216...27B}. The \textit{Chandra} X-ray luminosity of this cluster is $L_{[0.1-2.4\ \mathrm{keV}]} = (6.8 \pm 0.5) \times 10^{44}$ erg s$^{-1}$. The central temperature and morphology parameters suggest that it is a disturbed non-cool core cluster (Table \ref{tab:sample}).
The 325 MHz images are presented in Fig. \ref{fig:ULs1} \textit{top left panel}. No believable diffuse radio emission is visible in the low-resolution image. 
We derived an upper limit to the halo emission, which was found to be 5 mJy. 

\subsubsection{SPT-CL J0123-4821} \label{subsub:S0123-4821}
The SPT-CL J0123-4821 \citep{McDonald2013ApJ...774...23M} is a less massive ($M_{500}$ = $(4.5 \pm 0.9) \times 10^{14}$ $M_{\odot}$; \citealt{Bleem2015ApJS..216...27B}) cluster situated at the redshift $z = 0.655$ \citep{Bayliss2016ApJS..227....3B}. The \textit{Chandra} X-ray image shows a disturbed morphology and absence of bright core. The X-ray luminosity of this cluster was found to be $L_{[0.1-2.4\ \mathrm{keV}]} = (2.9 \pm 0.2) \times 10^{44}$ erg s$^{-1}$. The central temperature and morphology parameters suggest that it is a disturbed non-cool core cluster (Table \ref{tab:sample}).
The 325 MHz images are presented in Fig. \ref{fig:ULs1} \textit{top right panel}.
There is no diffuse emission visible in the low-resolution image contours. 
The radio halo upper limit for this cluster was placed to be 4 mJy.

\subsubsection{SPT-CL J0142-5032}
This is a massive cluster with $M_{500}$ = $(5.7 \pm 0.9) \times 10^{14}$ $M_{\odot}$; \citealt{Bleem2015ApJS..216...27B} and situated at the redshift $z = 0.6793$ \citep{Bayliss2016ApJS..227....3B}. The \textit{Chandra} X-ray image shows a disturbed morphology with luminosity of $L_{[0.1-2.4\ \mathrm{keV}]} = (4.3 \pm 0.5) \times 10^{44}$ erg s$^{-1}$. The central temperature and morphology parameters suggest that it is a disturbed non-cool core cluster (Table \ref{tab:sample}).
In Fig. \ref{fig:ULs1} \textit{middle left panel}, low-resolution image contours show that there is no diffuse emission from the ICM. 
Nothing was detected in the TGSS \citep{Intema2017A&A...598A..78I}, and SUMSS \citep{Bock1999AJ....117.1578B,Mauch2003MNRAS.342.1117M} survey, but the GLEAM \citep{Wayth2015PASA...32...25W,Hurley-Walker2017MNRAS.464.1146H} 170-231 MHz image cutout shows a faint emission at the position of the radio galaxies.
We estimated a radio halo upper limit for this cluster, which came out to be 10 mJy.

\subsubsection{SPT-CL J0212-4657}
The SPT-CL J0212-4657 \citep{McDonald2013ApJ...774...23M} is a less massive ($M_{500}$ = $(5.9 \pm 1.0) \times 10^{14}$ $M_{\odot}$; \citealt{Bleem2015ApJS..216...27B}) cluster situated at the redshift $z = 0.655$ \citep{Bleem2015ApJS..216...27B}. The \textit{Chandra} X-ray luminosity of this cluster is $L_{[0.1-2.4\ \mathrm{keV}]} = (5.4 \pm 0.6) \times 10^{44}$ erg s$^{-1}$. The central temperature and morphology parameters suggest that it is a disturbed non-cool core cluster (table \ref{tab:sample}).
The 325 MHz images are presented in Fig. \ref{fig:ULs1} \textit{middle right panel}. No believable diffuse radio emission is found in the low-resolution radio image (black contours). 
We searched for radio emission corresponding to the cluster region in TGSS \citep{Intema2017A&A...598A..78I}, SUMSS \citep{Bock1999AJ....117.1578B,Mauch2003MNRAS.342.1117M}, and found nothing. However, the GLEAM \citep{Wayth2015PASA...32...25W,Hurley-Walker2017MNRAS.464.1146H} survey detected a blob at the cluster position. Finally, because of the contamination from a nearby bright radio source, we could not place a halo upper limit to this cluster. 

\subsubsection{SPT-CL J0304-4921}
This cluster was first discovered in the Atacama Cosmology Telescope (ACT) SZE survey (ACT-CL J0304-4921) by \citet{Menanteau2010ApJ...723.1523M}. This is a massive cluster with $M_{500}$ = $(7.6 \pm 1.2) \times 10^{14}$ $M_{\odot}$; \citealt{Bleem2015ApJS..216...27B} and situated at the redshift $z = 0.392$ \citep{Sifon2013ApJ...772...25S}. The \textit{Chandra} X-ray image shows a regular morphology with bright central core. The X-ray luminosity of this cluster was found to be $L_{[0.1-2.4\ \mathrm{keV}]} = (6.0 \pm 0.3) \times 10^{44}$ erg s$^{-1}$. The central temperature and morphology parameters suggest that it is a relaxed strong cool core cluster (Table \ref{tab:sample}).
In Fig. \ref{fig:ULs1} \textit{bottom left panel}, low-resolution image contours show that there is no diffuse emission from the ICM. 
Because of this being a strong cool core cluster, we estimated a radio halo upper limit for this cluster corresponding to typical minihalo size i.e., 500 kpc, which was found to be 4 mJy.

\subsubsection{SPT-CL J0307-5042}
The SPT-CL J0307-5042 \citep{McDonald2013ApJ...774...23M} is a less massive ($M_{500}$ = $(5.3 \pm 0.9) \times 10^{14}$ $M_{\odot}$; \citealt{Bleem2015ApJS..216...27B}) cluster situated at the redshift $z = 0.55$ \citep{Bleem2015ApJS..216...27B}. The \textit{Chandra} X-ray luminosity of this cluster is $L_{[0.1-2.4\ \mathrm{keV}]} = (4.2 \pm 0.3) \times 10^{44}$ erg s$^{-1}$. The central temperature and morphology parameters suggest that it is a disturbed, non-cool-core cluster (Table \ref{tab:sample}).
The 325 MHz images are presented in Fig. \ref{fig:ULs1} \textit{bottom right panel}. No believable diffuse emission is found within the cluster region in the low-resolution radio image (black contours). However, an extension of the northern radio galaxy towards the east is visible outside the cluster region, which may be associated with the radio galaxy and not to the ICM, i.e., relic or fossil plasma.
We placed an upper limit to the radio halo for this cluster to be 5 mJy.

\subsubsection{SPT-CL J0411-4819}
This cluster was first discovered in the Planck Early Sunyaev-Zeldovich (ESZ) survey \citep{Planck2011A&A...536A...7P}. This is a massive cluster with $M_{500}$ = $(8.2 \pm 1.3) \times 10^{14}$ $M_{\odot}$; \citealt{Bleem2015ApJS..216...27B} and situated at the redshift $z = 0.424$ \citep{Bleem2015ApJS..216...27B}. The \textit{Chandra} X-ray image shows a disturbed morphology with a clear trace of merging activity. The X-ray luminosity of the system was found to be $L_{[0.1-2.4\ \mathrm{keV}]} = (7.4 \pm 0.3) \times 10^{44}$ erg s$^{-1}$. The central temperature and morphology parameters suggest that it is a disturbed non-cool core cluster (Table \ref{tab:sample}).
The 325 MHz images are presented in Fig. \ref{fig:ULs2} \textit{top left panel}. Although this cluster seems to be highly disturbed, no diffuse emission was found in the low-resolution image (black contours). 
We tried to place a reliable upper limit to the radio halo for this cluster, but contamination from a nearby bright radio source prevented the same.

\subsubsection{SPT-CL J0449-4901}
The SPT-CL J0449-4901 \citep{McDonald2013ApJ...774...23M} is a less massive ($M_{500}$ = $(4.6 \pm 0.9) \times 10^{14}$ $M_{\odot}$; \citealt{Bleem2015ApJS..216...27B}) cluster situated at the redshift $z = 0.792$ \citep{Bleem2015ApJS..216...27B}. The \textit{Chandra} X-ray image shows a disturbed morphology with X-ray luminosity $L_{[0.1-2.4\ \mathrm{keV}]} = (3.9 \pm 0.5) \times 10^{44}$ erg s$^{-1}$. The central temperature and morphology parameters suggest that it is a disturbed non-cool core cluster (Table \ref{tab:sample}).
In Fig. \ref{fig:ULs2} \textit{top right panel}, the low-resolution image shows an extended radio source with size of about $1.43\arcmin \times 2.58\arcmin$ or $0.64 \times 1.16$ Mpc (E-W $\times$ N-S), but the high-resolution image contours reveal it being a blended emission of multiple embedded radio galaxies. 
We estimated an upper limit to the radio halo for this cluster to be 4 mJy.

\subsubsection{SPT-CL J0456-5116}
This is a massive cluster with $M_{500}$ = $(5.1 \pm 0.9) \times 10^{14}$ $M_{\odot}$; \citealt{Bleem2015ApJS..216...27B} and situated at the redshift $z = 0.562$ \citep{Bleem2015ApJS..216...27B}. The \textit{Chandra} X-ray image shows a regular morphology with an absence of bright central core. The X-ray luminosity of this cluster was found to be $L_{[0.1-2.4\ \mathrm{keV}]} = (3.5 \pm 0.2) \times 10^{44}$ erg s$^{-1}$. However, the central temperature and morphology parameters suggest it being a disturbed non-cool core cluster (Table \ref{tab:sample}).
In Fig. \ref{fig:ULs2} \textit{bottom left panel}, low-resolution image contours show that there is no diffuse emission from the ICM. 
We estimated a radio halo upper limit for this cluster, which came out to be 10 mJy.

\subsubsection{SPT-CL J2301-4023}
The SPT-CL J2301-4023 \citep{McDonald2013ApJ...774...23M} or [LP96] 2259-4040 cluster was first discovered in optical survey conducted by \citet{Lidman1996AJ....112.2454L}. This is a less massive ($M_{500}$ = $(4.8 \pm 0.9) \times 10^{14}$ $M_{\odot}$; \citealt{Bleem2015ApJS..216...27B}) cluster situated at the redshift $z = 0.8349$ \citep{Bayliss2016ApJS..227....3B}. This is 
the most distant cluster in the analysed sample (15 clusters). The \textit{Chandra} X-ray luminosity of this cluster is $L_{[0.1-2.4\ \mathrm{keV}]} = (4.8 \pm 0.4) \times 10^{44}$ erg s$^{-1}$. The central temperature and morphology parameters suggest that it is a disturbed non-cool core cluster (Table \ref{tab:sample}).
The 325 MHz images are presented in Fig. \ref{fig:ULs2} \textit{bottom right panel}. No believable diffuse radio emission is found in the low-resolution image (black contours). 
A hint of the presence of the diffuse radio emission is found after subtraction of the radio galaxy flux density from the total emission, where the residual flux density was found to be $2.95 \pm 0.44$ mJy. However, further deeper observations are needed to reliably confirm the presence of any diffuse radio emission in this cluster. We have derived an upper limit to the halo in this cluster to be 7 mJy. 

\section{Discussion} \label{sec:discuss}
In this work, we present the radio observational results of the GSRHS sample of 15 galaxy clusters. Radio images were obtained at 325 MHz with $\sigma_{\mathrm{rms}} \sim$ 100 $\mu$Jy beam$^{-1}$ (Sect. \ref{sec:results}). Extended radio emission in the cluster ICM was found in 4 clusters, tentative detection of diffuse emission in 1 cluster and upper limits were estimated for non-detection in 8 clusters.

Below, we discuss the implications of our results in understanding the diffuse radio emission in galaxy clusters.

\subsection{Occurrence of radio halos} 
The fraction of radio halo detected in our sample is $f = 4/15 \sim 27\%$ and if we include 1 candidate halo, it becomes $f = 5/15 \sim 33\%$. It should be noted that these detection fractions are specific to the selected sample and will change according to the selection criteria. These occurrences are similar to what was previously reported by \citet{Kale2015A&A...579A..92K}, despite our sample being biased towards mergers. However, we note that an absolute comparison between these samples are not possible because of different selection criteria and the values presented here are only indicative.

\begin{figure*}
    \begin{tabular}{ccc}
    \includegraphics[width=0.64\columnwidth]{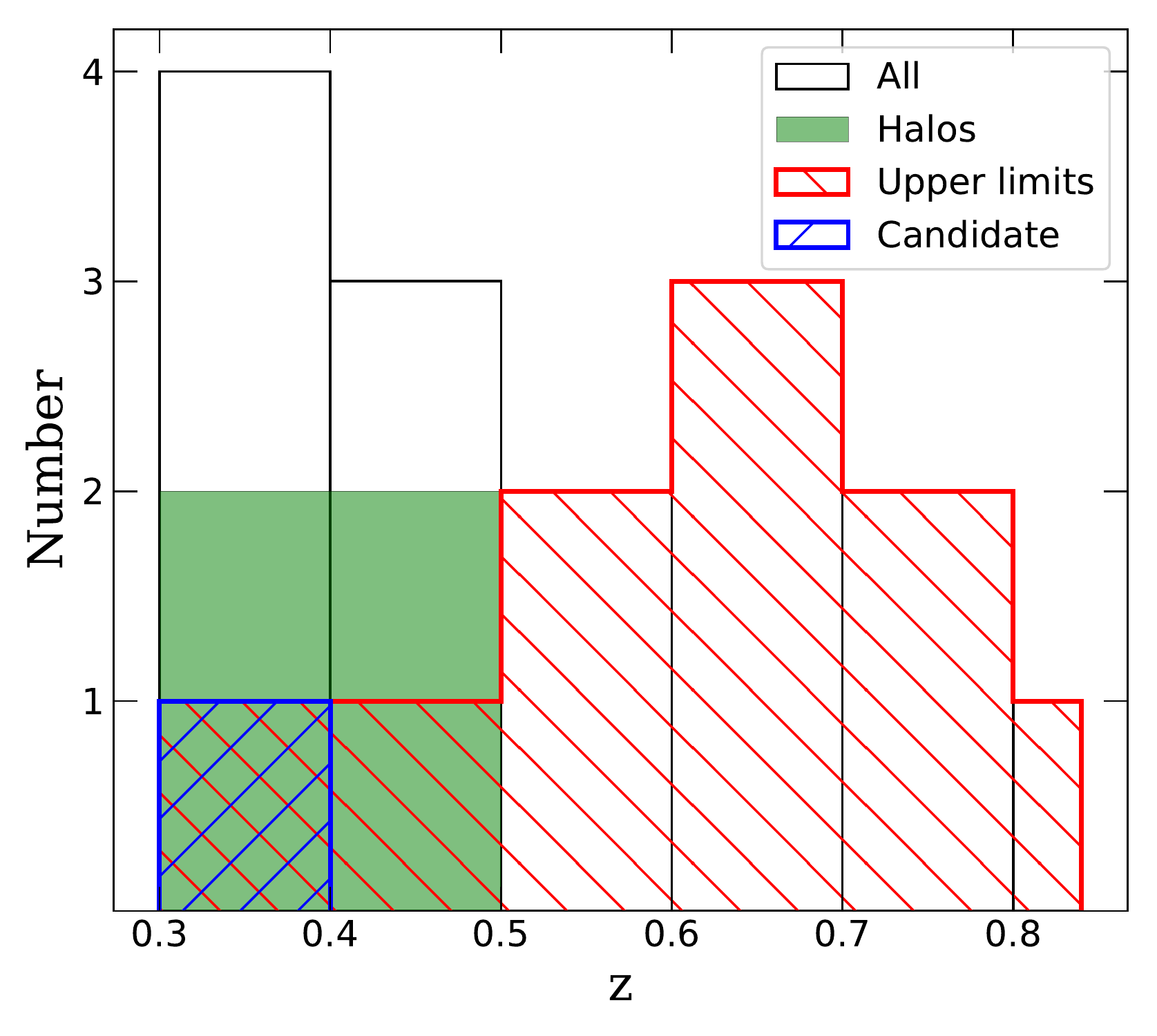} &
    \includegraphics[width=0.64\columnwidth]{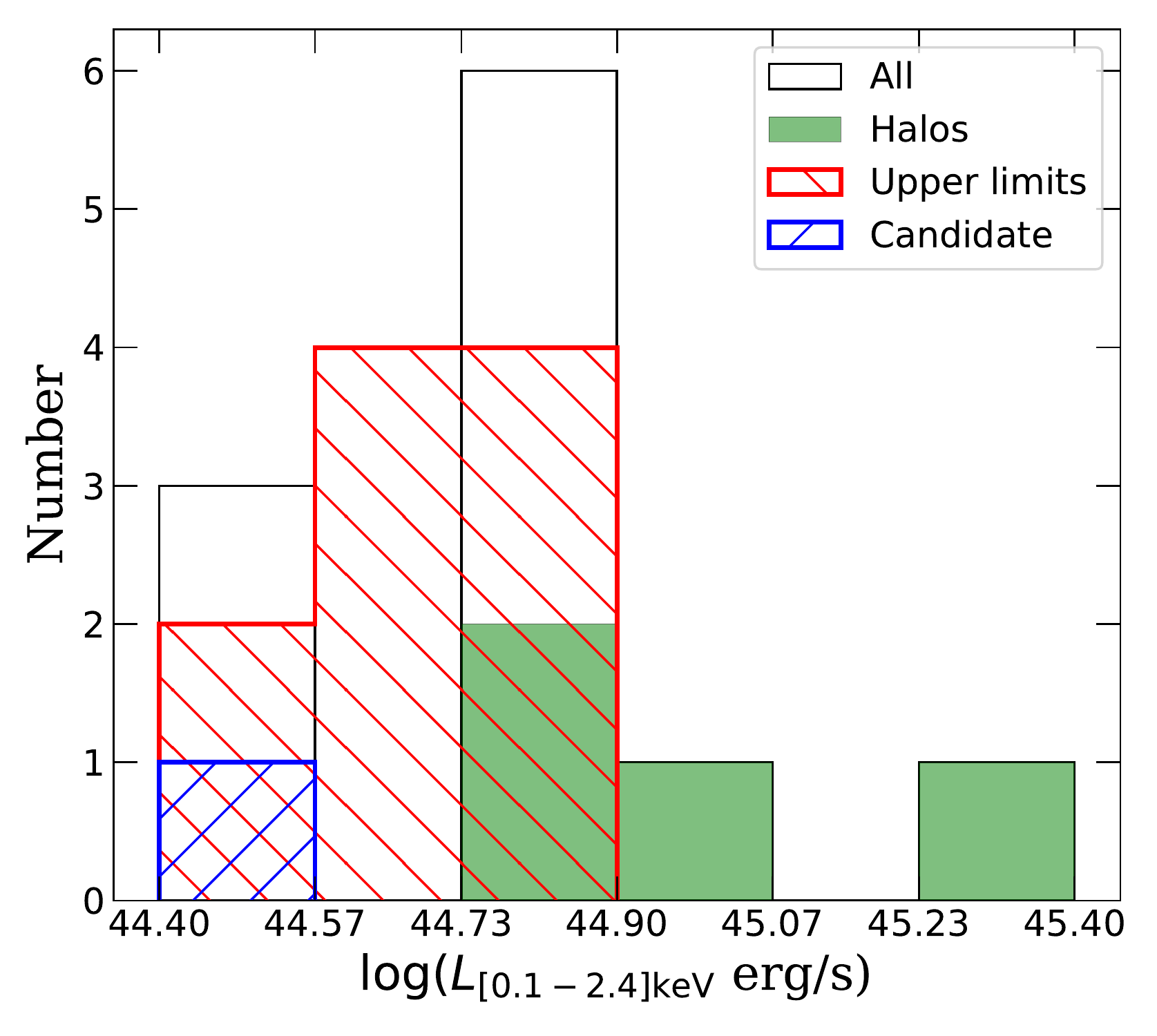} &
    \includegraphics[width=0.64\columnwidth]{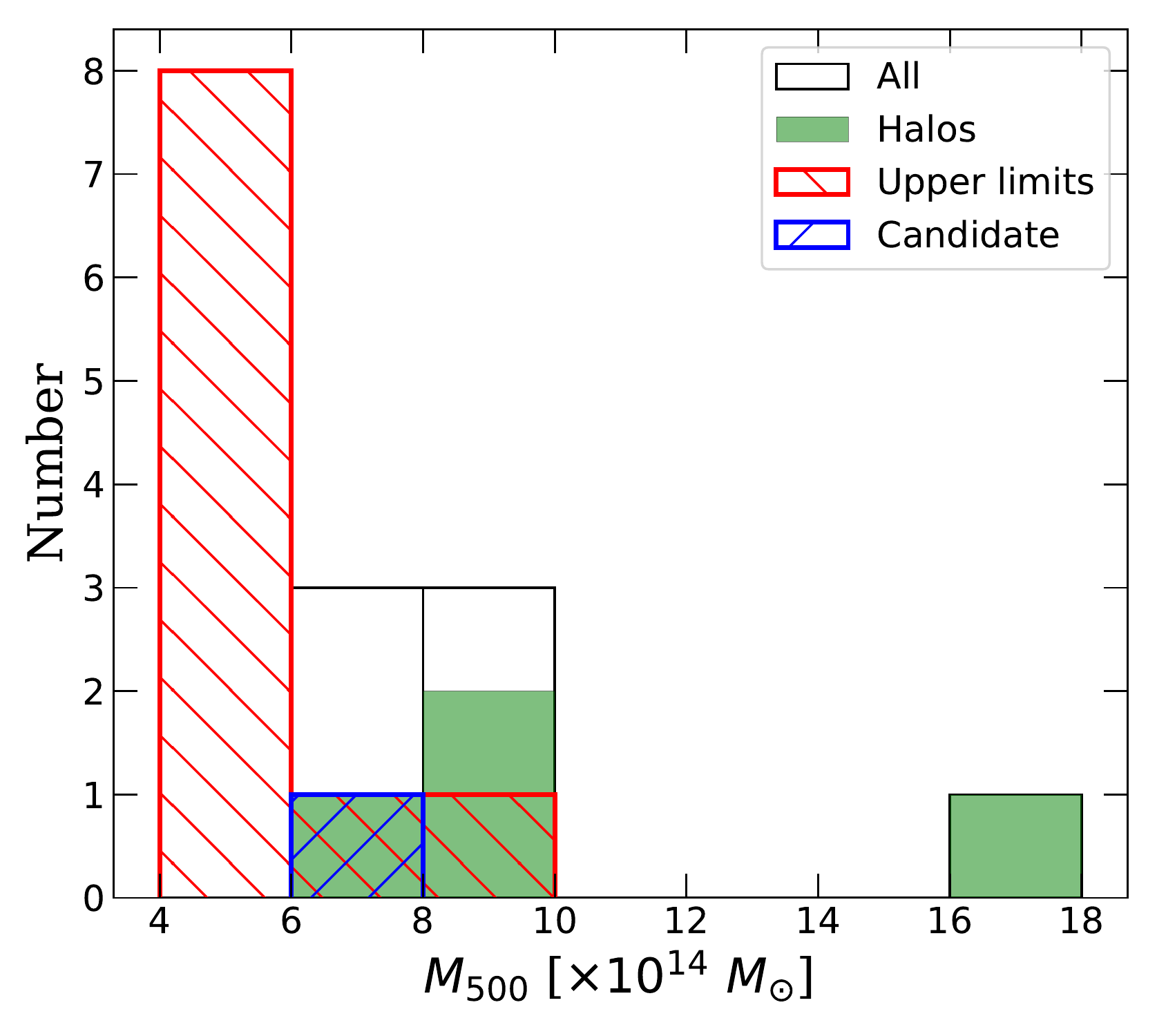}
    \end{tabular}
    \caption{Histograms show the distribution of halos, candidate halos and non-detections (upper limits) in the GSRHS sample across redshift (\textit{left panel}), X-ray luminosity (\textit{middle panel}), and mass (\textit{right panel}) of the clusters. The white histograms represent all clusters (All), the green histograms are radio halos (Halos), hatched histograms in blue and red colour are candidate halos (Candidates) and non-detection clusters (Upper limits), respectively.}
    \label{fig:z_Lx_M-dist}
\end{figure*}

In Fig. \ref{fig:z_Lx_M-dist}, we have illustrated with histogram plots the occurrence of radio halos across redshift, X-ray luminosity and mass. 
In Fig. \ref{fig:z_Lx_M-dist} \textit{left panel}, it can be seen that the detection of radio halos in each redshift bin is around 50\% from $z=0.3$ up to $z = 0.46$, and no detection beyond. The primary reason for this is the lack of sensitivity of our radio observations, and further deeper observations are needed to confirm whether there are any diffuse emission in these clusters or not. However, the lack of massive clusters in our sample at higher redshift, which is evident from Fig. \ref{fig:z_Lx_M-dist} \textit{right panel}, is also contributing to the lack of halos at higher redshift. 

In the \textit{middle panel} of Fig. \ref{fig:z_Lx_M-dist}, the radio halo distribution in X-ray luminosity shows that most of the clusters in our sample (13/15) are of low X-ray luminosity if we take the dividing X-ray luminosity to be $8 \times 10^{44}$ erg/s (or, 44.9 in log scale) following \citet{Venturi2008A&A...484..327V,Brunetti2009A&A...507..661B,Kale2015A&A...579A..92K}.
Also, in the \textit{right panel} of Fig. \ref{fig:z_Lx_M-dist} we see that most of the clusters with non-detection are in the $4-6 \times 10^{14}$ $M_{\odot}$ bin. So, although the clusters with non-detection in our sample are less X-ray luminous and less-massive, the real lack of sensitivity in our observations does not allow us to comment on the occurrence of halos in low-mass clusters.



\subsection{Distribution of halos in the \texorpdfstring{$P_{1.4} - L_\mathrm{X}$ and $P_{1.4} - M_{500}$\ }\ plane}
\begin{figure*}
    \begin{tabular}{cc}
    \includegraphics[width=\columnwidth]{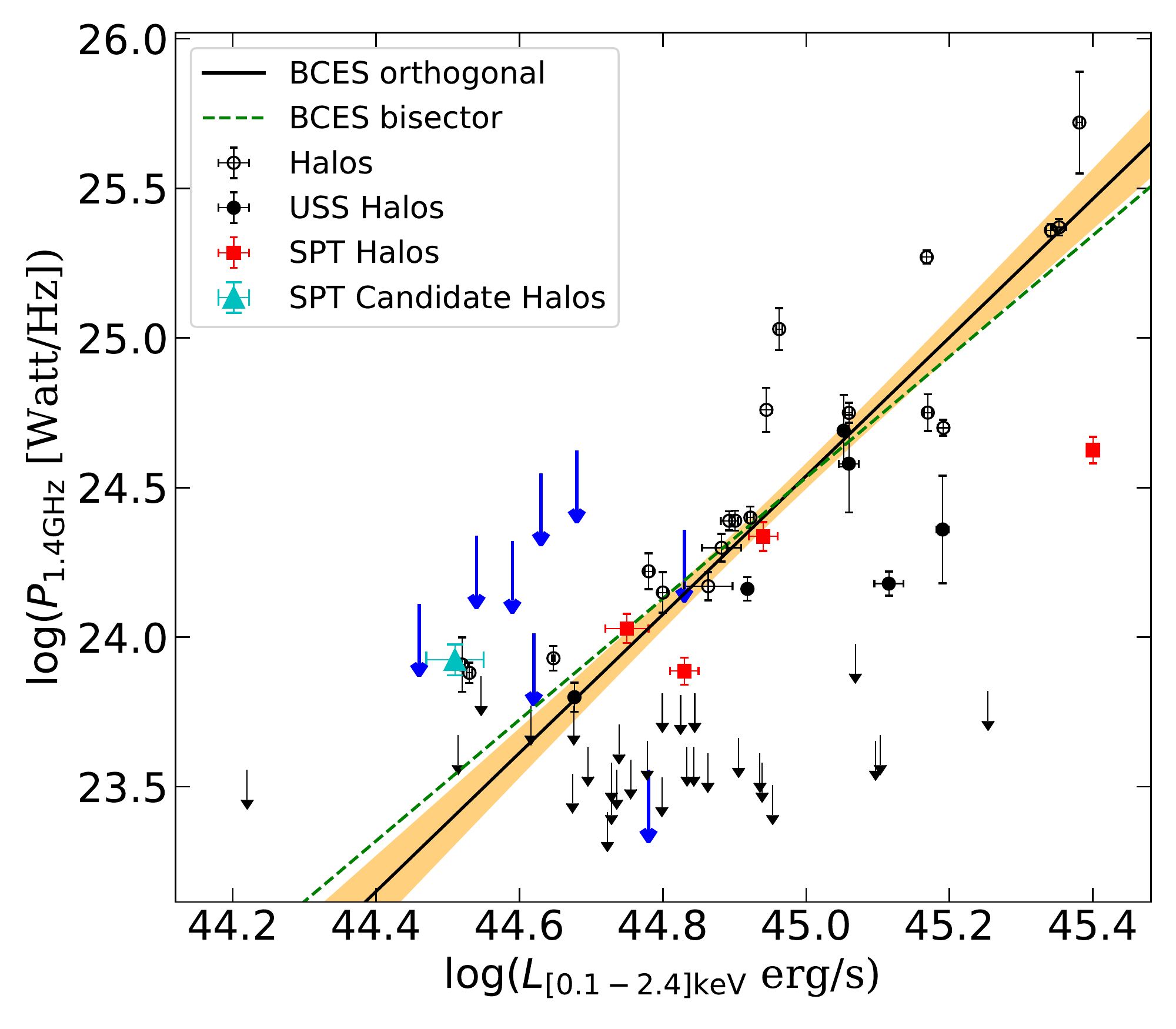} &
    \includegraphics[width=\columnwidth]{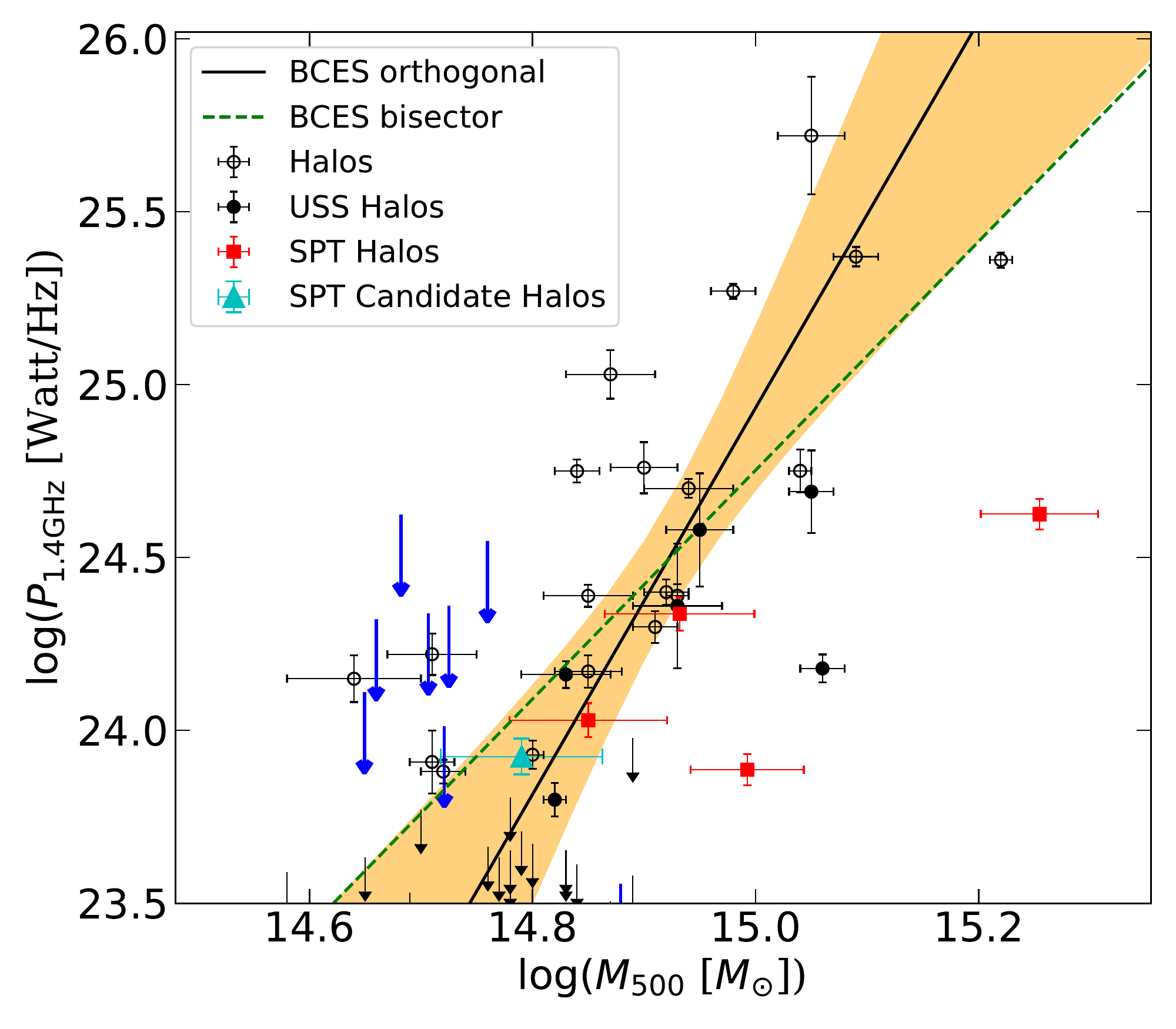}
    \end{tabular}
    \caption{The distribution of GSRHS halos, candidate and upper limits (blue arrows) along with literature halos, ultra-steep spectrum (USS) halos and upper limits (black arrows) taken from \citet{Cassano2013ApJ...777..141C} are plotted in the $P_{1.4} - L_\mathrm{X}$ (\textit{left panel}) and $P_{1.4} - L_{500}$ (\textit{right panel}). The mass information of the GSRHS clusters are taken from \citet{Bleem2015ApJS..216...27B}.The orange shaded region in both plots represents the 95\% confidence region of the best-fit relations.}
    \label{fig:corr}
\end{figure*}

It is established in the literature that there is a clear bimodal distribution of clusters with and without halos in both $P_{1.4} - L_\mathrm{X}$ and $P_{1.4} - M_{500}$ plane (e.g., \citealt{Cassano2013ApJ...777..141C,Kale2015A&A...579A..92K}). We have also plotted the distribution of our sample in both of these planes in Fig. \ref{fig:corr}. Although a proper comparison between samples in these correlations is possible only when the explored redshift range, as well as the sample selection criteria, is similar, halos in our sample (since they have similar redshifts as the literature sample) seem to lie within the range of scatter of the literature sample.

The $P_{1.4} - L_\mathrm{X}$ plot in Fig. \ref{fig:corr} \textit{left panel} shows that the clusters in our sample follow the observed correlation between radio and X-ray power. Both the halos and candidate are showing similar scatter as in literature halos. The only halo outlier is the SPT-CL J2248-4431 cluster, where the observed radio power is considerably lower than the correlation line. A multi-frequency analysis of this cluster was recently done by Rahaman et al. (submitted to MNRAS) where they argued the under luminous behaviour of this cluster being related to the merger being recent and is currently in switching-on stage and moving up towards the correlation. 
Additionally, most of the upper limits lie above the correlation lines, which indicates the possibility of hosting radio halo; merely the current data has insufficient sensitivity to detect them. So, they can not be considered with the non-halo group based on current observations. 
We derived a linear fit combining our halo sample (excluding the candidate halo) with the literature halo sample (including ultra-steep spectrum (USS) halos) using the BCES method \citep{Akritas1996ApJ...470..706A} of the form,
\begin{equation}
    \log(P_{1.4}) = A \times \log(L_\mathrm{X}) + B.
\end{equation}
The best fit parameters corresponding to the bisector and the orthogonal are $A=2.02 \pm 0.24$ and $B=-66.54 \pm 10.95$ and $A=2.32 \pm 0.30$ and $B=-79.65 \pm 13.63$, respectively, which are consistent with the previous studies (e.g., \citealt{Cassano2013ApJ...777..141C,Kale2015A&A...579A..92K}).

The plot of the samples in $P_{1.4} - M_{500}$ plane in Fig. \ref{fig:corr} \textit{right panel} also shows similar characteristics like the previous correlation. The distribution of our sample in this plane as well is similar to the literature sample with comparable scatter around the best-fit line. Here as well the only outlier is the massive cluster SPT-CL J2248-4431. Similar to the previous correlation, upper limits with higher redshifts lie above the best-fit line. Similarly, we derived a linear fit in this plane using BCES method \citep{Akritas1996ApJ...470..706A} of the form,
\begin{equation}
    \log(P_{1.4}) = A \times \log(M_{500}) + B
\end{equation}
using the same sample as previously. The best fit parameters corresponding to the bisector and the orthogonal came out to be $A=3.31 \pm 0.69$ and $B=-24.90 \pm 10.27$ and $A=5.61 \pm 1.78$ and $B=-59.26 \pm 26.50$, respectively, which are consistent with the previous study by \citet{Cassano2013ApJ...777..141C}.

\subsection{Dynamical states of the GSRHS sample}
Observational evidence of a close connection between diffuse radio sources like halos, relics and minihalos and the dynamical state of the clusters have already been reported in the literature in the past few decades. Halos and relics are connected with merging clusters, and minihalos are found in the relaxed clusters. 

\begin{figure}
    \centering
    \includegraphics[width=\columnwidth]{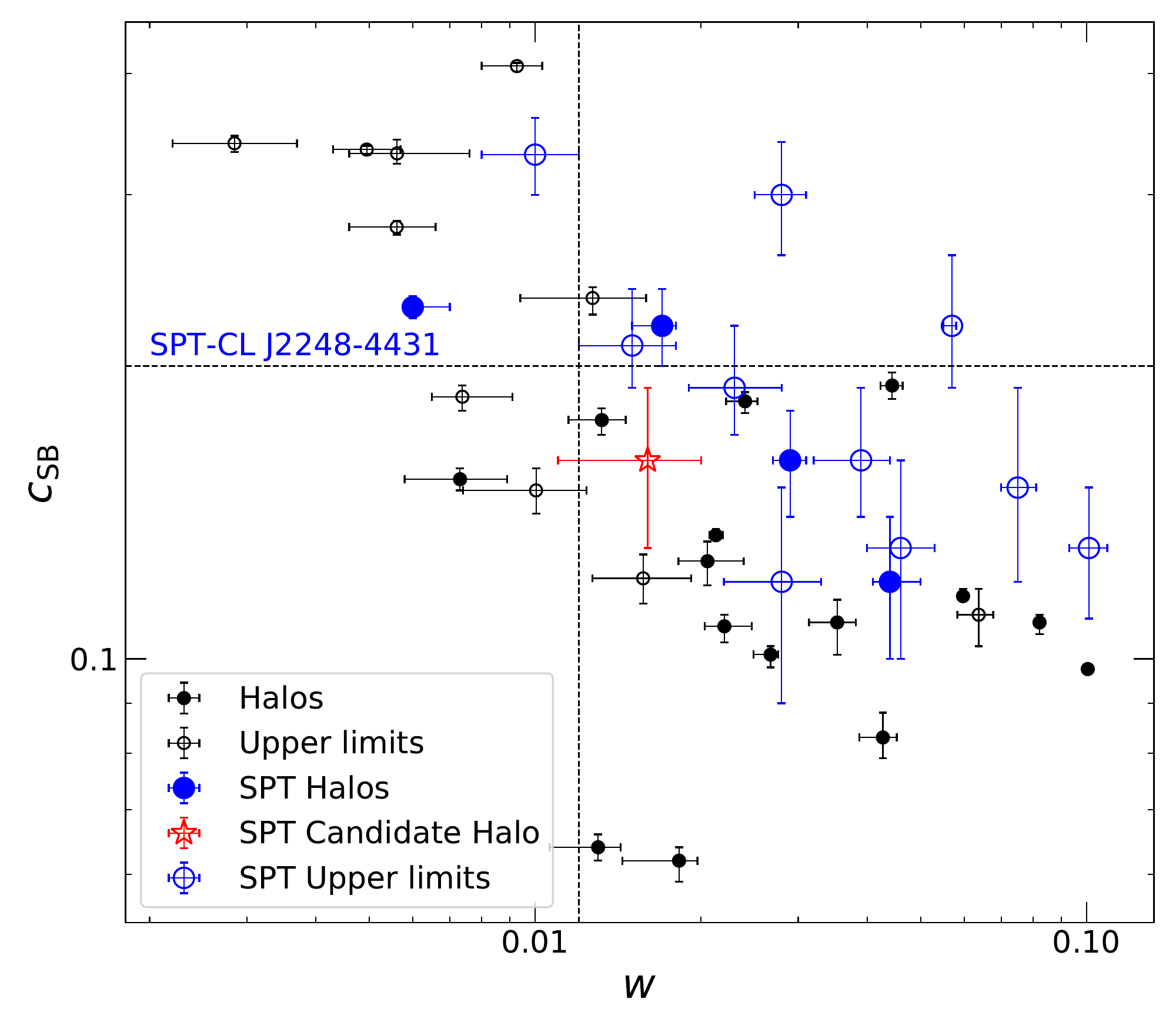}
    \caption{The distribution of GSRHS clusters along with literature clusters (halos and upper limits) reported in \citet{Cassano2016A&A...593A..81C} are plotted in the $c_\mathrm{SB} - w$ plane. We derived $c_\mathrm{SB}$ corresponding to the GSRHS clusters from the \textit{Chandra} data and $w$ information is taken from \citet{Nurgaliev2017ApJ...841...5N}.}
    \label{fig:c-w}
\end{figure}

Since the X-ray map traces the thermal distribution of a cluster ICM, it is widely used to estimate the dynamical state of the clusters. The dynamical state of a cluster is typically measured by estimating (i) the absence/presence of a bright core, e.g., surface brightness \lq\lq concentration parameter\rq\rq\ $c_\mathrm{SB}$ \citep{Santos2008A&A...483...35S} and (ii) disturbance in the ICM, e.g., \lq\lq centroid shift\rq\rq\ $w$ \citep{Mohr1993ApJ...413..492M}. In Fig. \ref{fig:c-w}, we have plotted these parameters in the $c_\mathrm{SB}-w$ plane corresponding to our sample along with the sample presented in \citet{Cassano2016A&A...593A..81C}. Following \citet{Santos2008A&A...483...35S} we derived $c_\mathrm{SB}$ for our sample from the \textit{Chandra} data and the $w$ information is taken from \citet{Nurgaliev2017ApJ...841...5N}. Following \citet{Cassano2010A&A...517A..10C}, we used $w=0.012$ and $c_\mathrm{SB}=0.2$ to separate merging clusters with radio halos and relaxed clusters without halos. 

Here, we see that despite being merging clusters, a lot of them does not show any diffuse emission.
But, unlike the literature sample where almost all are massive clusters, almost half of them in our sample are relatively low-mass clusters (Fig. \ref{fig:z_Lx_M-dist} \textit{right panel}). Specifically, most of the non-detections in Fig. \ref{fig:c-w} bottom right quadrant are in the first bin of the mass histogram (Fig. \ref{fig:z_Lx_M-dist} \textit{right panel}). 
However, we would like to point out that most of the clusters with the absence of halo in our sample are situated at higher redshifts, where the probability of halo detection is extremely low. Furthermore, the sensitivity of our observations is very poor to explore the possibility of hosting halos in these clusters. So, the reason for the absence of halo in our sample is not exactly similar to as \citet{Cuciti2015A&A...580A..97C}.
Nevertheless, since the possibility of detecting radio halo depends on several factors like mass of the system, merger energy injection fraction, magnetic field properties of the cluster, and merger stage (e.g, \citealt{Cassano2010ApJ...721L..82C,Donnert2013MNRAS.429.3564D,Brunetti2014IJMPD..2330007B}), sensitive low-frequency observations of merging clusters that do not show diffuse radio emission are crucial to clarify whether they host ultra-steep-spectrum halos or truly not host any diffuse radio emission at all.

\section{Summary and Conclusions} \label{sec:conclude}
The GSRHS consists of 15 clusters observed at 325 MHz frequency for a higher probability of radio halo detection. This study is one of the first systematic investigations of this kind in the redshift $z>0.3$ (e.g., \citealt{Knowles2019MNRAS.486.1332K, Giovannini2020A&A...640A.108G}). Although, the majority of clusters in this sample are less X-ray bright ($<8 \times 10^{44}$ erg/s) and half of the clusters are less massive ones ($4-6 \times 10^{14}$ $M_{\odot}$), this sample is biased towards merging clusters.

Radio images of halos, candidate, and upper limits are presented in Sect. \ref{sec:results}. We have discovered radio halo for the first time in SPT-CL J0013-4906 and SPT-CL J0304-4401 cluster. The presence of radio halo in 2 clusters of our sample, namely SPT-CL J2031-4037 and SPT-CL J2248-4431 were previously reported by \citet{Raja2020MNRAS.493L..28R} and \citet{Xie2020arXiv200104725X}, respectively. Also, SPT-CL J0348-4515 cluster is classified as a candidate for radio halo. Finally, we derived upper limits for the clusters where no diffuse emission was observed in SPT-CL J0014-4952, SPT-CL J0123-4821, SPT-CL J0142-5032, SPT-CL J0212-4657, SPT-CL J0304-4921, SPT-CL J0307-5042, SPT-CL J0411-4819, SPT-CL J0449-4901, SPT-CL J0456-5116 and SPT-CL J2301-4023. 
In addition, the halos and candidate follow the observed correlations in both $P_{1.4} - L_\mathrm{X}$ and $P_{1.4} - M_{500}$ plane with similar scatter as found in the literature data. Furthermore, the position of the derived upper limits above the correlation line indicates the possibility of future detection of diffuse radio emission with sensitive observations.
In conclusion, the study presented here is a pilot survey of high-redshift clusters which complements the previous GRHS and EGRHS in redshift distribution. 
Note that our current sample lacks sufficient massive clusters at high redshift. Future study of massive merging clusters with uniform redshift distribution is necessary to investigate the redshift dependence of the radio halos. Furthermore, a systematic investigation of less-massive, dynamically disturbed clusters at low redshift is crucial in understanding the nature of radio halos in general.
Finally, future sensitive low-frequency surveys with uGMRT, JVLA, MeerKAT, LOFAR, MWA and the upcoming SKA will be the key in studying the unexplored aspects of galaxy clusters to understand the origin and evolution of different diffuse radio structures found in them.

\section*{Acknowledgements}

We would like to thank IIT Indore for providing necessary computing facilities for data analysis. We thank the staff of GMRT, who made these observations possible. GMRT is run by the National Centre for Radio Astrophysics of the Tata Institute of Fundamental Research. This work has made use of data from the Chandra Data Archive. This research is supported by DST-SERB, through ECR/2017/001296 grant awarded to AD. MR would like to thank DST for INSPIRE fellowship program for financial support (IF160343). This research has made use of the SIMBAD database,
operated at CDS, Strasbourg, France \citep{Wenger2000A&AS..143....9W}. This research made use of Astropy,\footnote{http://www.astropy.org} a community-developed core Python package for Astronomy \citep{AstropyCollaboration2013A&A...558A..33A,AstropyCollaboration2018AJ....156..123A}. This research made use of Matplotlib \citep{matplotlib} and APLpy \citep{aplpy}, open-source plotting packages for Python.

\section{Data Availability}
All the radio data used in this study are available in the GMRT Online Archive (\url{https://naps.ncra.tifr.res.in/goa/data/search}) with proposal code 26\_024 and 27\_026.



\bibliographystyle{mnras}
\bibliography{bibliography} 








\bsp	
\label{lastpage}
\end{document}